\documentclass[10pt]{amsart}
\allowdisplaybreaks[4]
\usepackage{amssymb,color}
\usepackage{amsmath,amssymb,graphicx}
\usepackage{accents}
\usepackage{bbm}
\usepackage{lipsum}
\usepackage{array}
\usepackage{multirow}

\usepackage{tabularx,ragged2e,booktabs,caption}
\usepackage{here}
\usepackage[toc,page]{appendix}

\usepackage[colorlinks=true, citecolor=blue, linkcolor=blue, urlcolor=blue ]{hyperref}

\usepackage{amsthm}
\usepackage{subcaption}

\def\FRE{\mbox{Fr\'{e}chet} }

\definecolor{c10}{rgb}{0.,0.7,0.}
\definecolor{c20}{rgb}{0.,0.,0.}
\definecolor{c30}{rgb}{0.,0.,0.}
\definecolor{c40}{rgb}{0,0,0}
\definecolor{c50}{rgb}{0,0,0}
\definecolor{c60}{rgb}{0,0.,0.}
\definecolor{c30}{rgb}{0.9,0.5,0.5}

\newcommand{\ABs}[1]{ \biggl \lvert #1 \biggr \rvert}

\newcommand{\EE}[1]{\mathbb{E}\left\{ #1 \right\}}

\newcommand{\pk}[1]{\mathbb{P} \left\{#1 \right\} }

\newcommand{\R}{\mathbb{R}}

\newcommand{\inr}{\in \R}

\newcommand{\ldot}{,\ldots,}

\newcommand{\limit}[1]{\lim_{#1 \to   \infty}}

\newcommand{\BQN}{\begin{eqnarray}}
\newcommand{\EQN}{\end{eqnarray}}
\newcommand{\BQNY}{\begin{eqnarray*}}
\newcommand{\EQNY}{\end{eqnarray*}}

\def\CE#1{\textcolor{c20}{#1}}
\def\CE#1{{#1}}
\def\green#1{\textcolor{c20}{#1}}
\def\GM#1{\textcolor{c10}{#1}}
\def\GM#1{{#1}}
\def\ehm#1{\textcolor{c10}{#1}}
\def\ehm#1{{#1}}
\def\MG#1{\textcolor{c10}{#1}}
\def\MG#1{{#1}}

\def\bqny#1{{ \begin{eqnarray*} #1 \end{eqnarray*}}}
\def\bqn#1{{ \begin{eqnarray} #1 \end{eqnarray}}}

\newcommand{\BS}{\begin{sat}}
\newcommand{\ES}{\end{sat}}
\newcommand{\BT}{\begin{theo}}
\newcommand{\ET}{\end{theo}}
\newcommand{\BK}{\begin{korr}}
\newcommand{\EK}{\end{korr}}

\newcommand{\BD}{\begin{de}}
\newcommand{\ED}{\end{de}}

\newcommand{\BIT}{\begin{itemize}}
\newcommand{\EIT}{\end{itemize}}
\newcommand{\BDI}{\begin{description}}
\newcommand{\EDI}{\end{description}}

\newcommand{\BRM}{\begin{remarks}}
\newcommand{\ERM}{\end{remarks}}

\newcommand{\BEL}{\begin{lem}}
\newcommand{\EEL}{\end{lem}}

\newtheorem{theo}{Theorem}[section]
\newtheorem{sat}[theo]{Proposition}
\newtheorem{de}[theo]{Definition}
\newtheorem{lem}[theo]{Lemma}

\newtheorem{korr}[theo]{Corollary}

\newtheorem{remarks}[theo]{Remarks}

\newcommand{\proofprop}[1]{\textbf{Proof of Proposition} \ref{#1}}

\newcommand{\COM}[1]{}

\newcommand{\QED}{\hfill $\Box$ \\}

\def\IF{\infty}

\newcommand{\toprob}{ \stackrel{p}{\to}}

\def\ehe#1{\textcolor{c50}{#1}}
\def\ehe#1{#1}
\def\ebe#1{\textcolor{c50}{#1}}
\def\ebe#1{#1}

\def\ece#1{\textcolor{c50}{#1}}
\def\ece#1{#1}

\def\IF{\infty}

\begin{document}

\title{\GM{On Some New  Dependence Models derived from Multivariate Collective Models in Insurance Applications}}

\author{Enkelejd Hashorva}
\address{Enkelejd Hashorva, Department of Actuarial Science, University of Lausanne, UNIL-Dorigny 1015 Lausanne, Switzerland}
\email{enkelejd.hashorva@unil.ch}

\author{Gildas Ratovomirija}
\address{Gildas Ratovomirija, Department of Actuarial Science, University of Lausanne, UNIL-Dorigny 1015 Lausanne, Switzerland, 
and 	Vaudoise Assurances, Place de Milan CP 120, 1001 Lausanne,         Switzerland}
\email{gildas.ratovomirija@unil.ch / gra@vaudoise.ch}

\author{Maissa Tamraz}
\address{Maissa Tamraz, Department of Actuarial Science, University of Lausanne, UNIL-Dorigny 1015 Lausanne, Switzerland}
\email{maissa.tamraz@unil.ch}

\bigskip

\date{\today}
 \maketitle

\bigskip
{\bf Abstract:} 
Consider two different portfolios which have claims triggered by the same events. Their corresponding 
collective model over a fixed time period is given in terms of individual claim sizes $(X_i,Y_i), i\ge 1$ and a claim counting random variable $N$. In this paper we are concerned with the joint distribution function $F$ of the \ece{largest claim sizes} $(X_{N:N}, Y_{N:N})$.  By allowing $N$ to depend on some parameter, say $\theta$, then $F=F(\theta)$ is for various choices of $N$ a tractable parametric family of bivariate distribution functions. 
\GM{
 We investigate both distributional and extremal properties of  $(X_{N:N}, Y_{N:N})$.
 Furthermore, we present several  applications of the implied parametric models to some data from the literature and a new data set from a Swiss insurance company \footnote{Data set can be downloaded here 
\url{http://dx.doi.org/10.13140/RG.2.1.3082.9203}}.   }
\\



\section{Introduction}
\GM{ Modelling the dependence structure between insurance risks  is one of the main tasks of actuaries.  For instance, the determination of a risk capital in the risk management framework needed to cover unexpected losses of an  insurance portfolio and the allocation of the latter to each line of business is of importance when choosing the best model of dependence for multivariate insurance risks.
As discussed in Nelsen (1999), copulas are a popular multivariate distribution when modelling  the dependency between insurance risks as they separate the marginals from the dependence structure, see Embrechts (2009), Genest et al. (2009) and references therein.
With motivation from Zhang and Lin (2016), in this contribution we propose a flexible family of copulas  derived from  the joint distribution of the largest claim sizes of two insurance portfolios. 
}
\\
Next, in order to introduce our model, we consider the \ebe{classical} collective model over a fixed time period of two insurance portfolios with 
$(X_i,Y_i)$ \ebe{modelling} the $i$th claim \ebe{sizes}  of both portfolios and $N$ the total number of such claims. If $N=0$, then there are no claims, so the largest claims in both portfolios are equal to 0. When $N\ge 1,$ then $(X_{N:N},Y_{N:N})$ denotes the maximal claim amounts in both portfolios. \ece{Commonly, claim sizes are assumed to be positive, however here} 
we shall simply assume that $(X_i,Y_i), i\ge 1 $ are independent with common distribution function (df) $G$ and $N$ is independent of everything else. 
 Such a model is common for proportional reinsurance. In that case $Y_i= c X_i$ with $c$ being a positive constant. Another instance is if $X_i$'s model claim sizes and $Y_i$'s model the expenses related to the settlement of $X_i$'s, see Denuit et al. (2005) for statistical treatments and further applications. The df of $(X_{N:N},Y_{N:N})$ denoted by  $F^*$ is given by
\BQN\label{intro}
 F^*( x,y)= 
 L_N(-\ln G(x,y)), \quad x,y \ebe{\ge}  0, 
 \EQN
with $L_N$ the Laplace transform of $N$.  Clearly, $\ebe{F^*}$ is a mixture \ebe{df given by}
$$ 
F^*(x,y)= \pk{N=0}+ \pk{N\ge 1} F(x,y), \quad x,y \ebe{\ge} 0,
$$
where 
\BQN\label{intro0}
 F(x,y)=  L_\Lambda(-\ln G(x,y)), \quad x,y\ge 0,
\EQN
with  $\Lambda=N \lvert N \ge 1$ and $L_\Lambda$ its Laplace transform. \\
Since both distributional and asymptotic properties of $F^*$ \ebe{can be} easily derived from those of $F$, in this paper 
we shall focus on $F$ assuming throughout that $\Lambda\ge 1$ is an integer-valued random variable. \\
 When \ebe{the df} $G$ is a product distribution, $F$ above corresponds to the frailty model,  see e.g., Denuit et al. (2005), 
 whereas the special case 
that $\Lambda$ is a shifted geometric random variable is dealt with in  Zhang and Lin (2016). 
We mention three tractable cases for $\Lambda$:\\
{\bf Model A}: In  Zhang and Lin (2016), $\Lambda$ is assumed to have a shifted Geometric distribution with parameter $\theta \in (0,1)$ which leads to 
\BQN\label{exA} 
 F(x,y)=  \frac{\theta G(x,y)}{1- (1- \theta) G(x,y)}, \quad \ebe{x,y\ge 0}.
 \EQN
{\bf Model B}:  $\Lambda$ has a shifted Poisson distribution with parameter $\theta >0$, i.e.,  $\Lambda= 1+ K$ with $K$ being a Poisson random variable with mean $\theta>0$, which implies 
\BQN \label{exB}
 F(x,y)=  G(x,y) e^{ - \theta[1- G(x,y)]} , \quad \ebe{x,y\ge 0}.
 \EQN
{\bf Model C}:   $\Lambda$ has a truncated Poisson distribution  with $\pk{\Lambda = k}= e^{- \theta} \theta ^k /(k! (1- e^{-\theta})), k\ge 1$ and thus 
\BQN \label{exC}
 F(x,y)=  \frac{e^{-\theta} }{1-e^{-\theta}} [ e^{  \theta G(x,y)} -1], \quad \ebe{x,y\ge 0}. 
 \EQN
\ehm{Since the distributions $F$ and their copulas are indexed by an unknown parameter $\theta$, the new mixture copula family has several interesting properties. In particular, it allows to model highly dependent insurance risks and therefore our model is suitable for numerous insurance applications including risk aggregation, capital allocation and reinsurance premium calculations.}
 
In this contribution we \ebe{investigate} first \ehm{the basic distributional and extremal} properties of $F$ for general $\Lambda$. \ece{As it will be shown in Section 3}, interestingly the extremal properties of $F$ are similar to those of $G$. 

With some motivation from Zhang et al. (2016), which investigates {\bf Model A} \ebe{and its applications}, 
in this paper, we shall discuss \ebe{parameter estimation and Monte Carlo simulations  for parametric families of bivariate \ece{df's} induced by $F$.} In particular, we apply our \ebe{results to actuarial modelling  of} concrete data sets from actuarial literature. Moreover we shall consider the implications of our findings for a new real data set from a Swiss insurance company. In several cases {\bf Model B} and {\bf Model C} give both satisfactory fit to the data. For the case of Loss and ALAE data set we model further \CE{the  stop loss and} the excess of loss  reinsurance premium. 
\GM{ One of the applications of the joint distribution of the largest claims $(X_{N:N}, Y_{N:N})$ of two insurance portfolios is the analysis of the impact of their sum on the risk profile of the  portfolios.
  Over the last decades, many contributions have been devoted on the  study of the  influence of the largest claims on aggregate claims, see e.g., Peng (2014) , Asimit and Chen (2015)  for an overview of existing contributions on the topic. 
  This analysis is important when designing risk management  and  reinsurance strategies especially in non proportional reinsurance. 
  Ammeter (1964)  is one of the first contribution  which  addressed  the impact of the largest claim $X_{N:N}$  on the moments of the total loss of an insurance portfolio $\sum_{i=1}^N X_i$, \ehm{see also Asimit and Chen (2015)  for recent results.} In this paper we demonstrate by simulation the influence of the sum of the largest claims observed in  two insurance portfolios  $ X_{N:N} + Y_{N:N}$  on the distribution of  $S_N$. Moreover, using the covariance capital allocation principle we quantify the impact of  $X_{N:N}$ and $Y_{N:N}$ on the total loss $S_N$.}
\GM{The paper is organised as follows.  We discuss next some basic distributional properties of $F
$.
An investigation  of 
the coefficient of upper tail dependence and the max-domain  of attractions of $F$ is presented in Section 3.
Section 4 is dedicated to parameter estimation and \ebe{Monte Carlo} simulation with special focus on the cases covered by 
 {\bf Model A-C} above. 
We present three applications to concrete insurance data set in Section 5. 
All the proofs are relegated to Appendix.
}

\COM{
Organisation of the rest of the paper. 
 We discuss next some basic distributional properties of $F
$.  Section 3 is dedicated to parameter estimation and \ebe{Monte Carlo} simulation with special focus on the cases covered by 
 {\bf Model A-C} above. 
We present four applications to concrete insurance data set in Section 4 followed by an investigation  of 
the coefficient of upper tail dependence and the max-domain  of attractions of $F$.  
\
}
\section{Basic  Properties of $F$} 
Let $G$ denote the df of $(X_1,Y_1)$ and write $G_1,G_2$ for its marginal \ebe{df's}. 
 Suppose that $G_i$'s are continuous and thus the copula \ehe{$Q$} of $G$ is unique.  
 For $\Lambda = N \lvert N\ge 1$,  we have that the marginal df's of $F$ are 
$$ F_i(x)= L_\Lambda(-\ln G_i(x_i)), \quad i=1,2, x\inr.$$
Hence, the generalised inverse of $F_i$ is 
$$F_i^{-1}(q)= G_i^{-1}( e^{-  L_\Lambda^{-1}(q)}), \quad q\in (0,1),$$
 where $G_i^{-1}$ is the generalised inverse of $G_i, i\le d$. 
Consequently, since the continuity of $G_i$'s implies that of $F_i$'s, the unique copula \ece{$C$} of $F$ is given by 
\BQN\label{CopulaF}
 C( u_1,u_2)&=&  F( F_1^{-1}(u_1), F_2^{-1}(u_2) ) \notag \\
&=&  L_\Lambda\bigl(-\ln G(G_1^{-1}(v_1) ,G_2^{-1}(v_2) )\bigr)\notag \\
&=&  L_\Lambda\bigl(-\ln Q(v_1 ,v_2)\bigr), \quad  \quad u_1,u_2\in [0,1],
\EQN
where we set 
$$ v_i=  e^{-  L_\Lambda^{-1}(u_i)}.$$
\GM{
\BRM \label{rem:copulaConv}
The df of the bivariate copula in \eqref{CopulaF}  can be extended to the multivariate case. Let  $X_j^{(i)} $ be the $j$-th claim sizes  of the portfolio $i, i=1,\ldots,d$ and $ j=1,\ldots, N$. Thus, the df of 
 $(X_{N:N}^{(1)}, \ldots, X_{N:N}^{(d)} )$ is given by 
 \BQNY 
 		F( z_1 ,\ldots, z_d)
			&=&  L_\Lambda\bigl(-\ln G(z_1 ,\ldots, z_d)\bigr), \quad z_1 \ldot z_d \inr, 
\EQNY 
where $G$ is the df of $(X_{1}^{(1)}, \ldots, X_{1}^{(d)} )$.
Similarly to the bivariate case one may express the copula of $F$ as follows
\BQNY 
 C( u_1 ,\ldots, u_d)
&=&  L_\Lambda\bigl(-\ln Q(v_1 ,\ldots, v_d)\bigr), \quad  \quad u_1, \ldots, u_d \in [0,1],
\EQNY 
where $Q$ is the copula of $G$. Without loss of generality, we present  in the rest of the paper the results for the bivariate case.
\ERM
}
Next, if $G$ has a probability density function (pdf) $g$, then $Q$ has a pdf $q$ given by 
$$ q(u_1,u_2)=  \frac{ g(G_1^{-1}(u_1), G_2^{-1}(u_2))}{ g_1(G_1^{-1}(u_1)) g_2(G_2^{-1}(u_2))}, \quad u_1,u_2 \in [0,1],$$
with $g_1,g_2$ the marginal pdf's. Consequently, the \ebe{pdf} $c$ of $C$ is given by (set 
$t=-\ln Q (v_1,v_2)$)
\BQN \label{eq:Copula_PDFF}
c(u_1,u_2) 
&=&\frac{\frac{\partial v_1}{\partial u_1}\frac{\partial v_2}{\partial u_2}}{Q^2(v_1,v_2)}
\biggl(\bigl(L'_\Lambda(t)+L''_\Lambda(t)\bigr)\frac{\partial Q(v_1,v_2)}{\partial v_1}\frac{\partial Q(v_1,v_2)}{\partial v_2}  -L'_\Lambda(t) Q(v_1,v_2)q(v_1,v_2) \biggr),
\EQN
where $L_\Lambda'(s)= -  \Lambda e^{- s \Lambda}$ and $L_\Lambda''(s)= 
\EE{ \Lambda^2 e^{- s \Lambda}}$.
The explicit form of $c$ for tractable copulas $Q$ and Laplace transform \ehe{$L_\Lambda$} is useful 
for the pseudo-likelihood method of 
parameter estimation treated in Section 4. 

To this end, we briefly discuss the correlation order and its implication for the dependence exhibited by $F$. 
Clearly, for any $x,y$ non-negative 
$$ F(x,y)\le   G(x,y).
$$
\ebe{Consequently}, in view of the correlation order, see e.g., Denuit et al. (2005) we have that Kendall's tau $\tau(X_{\Lambda:\Lambda}, Y_{\Lambda: \Lambda})$, Spearman's rank correlation $\rho_S(X_{\Lambda:\Lambda}, Y_{\Lambda: \Lambda})$ and 
the correlation coefficient $\rho(X_{\Lambda:\Lambda}, Y_{\Lambda: \Lambda})$ (when it is defined) are 
bounded by the same dependence measures calculated to $(X_1,Y_1)$ with df $G$, respectively.\\
Moreover, if  $\EE{\Lambda}< \IF$, then by applying Jensen's inequality (recall $\Lambda \ge 1$ almost surely)  for any $x,y$ non-negative 
\BQN \label{eq: corr_Order}
G^{a}(x,y)\le  G^{ \EE{\Lambda}}(x,y)= e^{ \EE{\Lambda}\ln G(x,y)}  \le
 \EE{e^{ \Lambda \ln G(x,y)}}    \le   F(x,y), 
 \EQN
with  $a$ the smallest integer larger than $\EE{\Lambda}$. Since $G^a$ is a df, say of $(S,T)$, then 
again the correlation order implies that $\tau(X_{\Lambda:\Lambda}, Y_{\Lambda: \Lambda}) \ge \tau(S,T)$, 
and similar bounds hold  for Spearman's rank correlation and the correlation coefficient. In the following 
we shall write also $\tau(C)$ and $\tau(Q)$ (if $a=1)$ instead of $\tau(X_{\Lambda:\Lambda}, Y_{\Lambda: \Lambda})$ and 
$ \tau(S,T)$, respectively. \CE{Similarly, we denote $\rho_S(C)$ and $\rho_S(Q)$ instead of $\rho_S(X_{\Lambda:\Lambda}, Y_{\Lambda: \Lambda})$ and 
$ \rho_S(S,T)$, respectively. } \\

\section{Extremal Properties of $F$ }
\ehm{In this section, we investigate the extremal properties of $F$ and its copula.} Assume that $\Lambda= \Lambda_n$ \ehe{depends on $n$ and} write $C_n$ instead of $C$. Suppose for simplicity  that $\EE{\Lambda_n}=n$ and 
$G$ has unit \FRE margins. Assume 
additionally the following convergence in probability  
\bqn{ \label{Z}
	\frac{\Lambda_n}{n} \toprob 1, \quad n\to \IF.
} 
The above conditions can be easily verified in concrete examples, in particular it holds if $\Lambda_n=n$ almost surely.\\
In order to understand the dependence of $C_n$, we can calculate Kendall's tau $\tau(C_n)$ as $n\to \IF$.  \COM{The \ehe{simulation results} suggest that 
	$\limit{n} \tau(C_n)=0$ }
\GM{ For instance, as shown in the simulation results in Table \ref{table: depMes}, if the copula $Q$ of $G$ has a coefficient of upper tail dependence $\mu_Q=0$, then $\limit{n} \tau(C_n)=0$. }
Note that by definition if $\mu_Q$ exists, then it is calculated by 
\BQN \label{muCQ}
\mu_Q&=& 2- \lim_{u \downarrow 0} u^{-1}[ 1- Q(1- u, 1-u)] \in [0,1].
\EQN
The following result establishes the convergence of both Kendall's tau for $C_n$ and Spearman's rank correlation $\rho_S(C_n)$ to the corresponding measures of dependence with respect to an extreme value copula \CE{$Q_A$} which \ece{approximates} $Q$, i.e., 
\BQN\label{copA}
\limit{n} \sup_{u_1,u_2 \in [0,1]}\ABs{ (Q(u_1^{1/n},u_2^{1/n}))^n -  \green{Q}_A(u_1,u_2)}=0,
\EQN
where 
\BQN\label{2a}
\green{Q}_A(u_1,u_2)= (u_1 u_2)^{A( y/ (x+y))} , \quad x= \ln u_1, y= \ln u_2
\EQN
for any $(u_1, u_2) \in (0,1]^2 \setminus {(1,1)},$  with $A: [0,1] \to [1/2, 1]$ a  convex function which satisfies 
\BQN\label{2}   
\max (t, 1- t) \le A(t) \le  1, \quad \forall t\in [0,1].
\EQN
In the literature, see e.g., Falk et al. (2010); Molchanov (2008); B{\"u}cher and Segers (2014); Aulbach et al. (2015), $A$ is referred to as the Pickands dependence function. 

\BS \label{tauHok}
If the copula $Q$ satisfies \eqref{copA} and \ece{further} \eqref{Z} holds, then 
\bqn{\label{tau}
	\limit{n}\tau(C_n)= \tau ( \CE{Q_A } ), \quad 
	\limit{n}\rho_S(C_n)= \rho_S(\CE{Q_A }).
}
\ES
If 
\CE{$Q_A$} is different from the independence copula, and therefore $\CE{A}(t)< 1$ for any $t\in (0,1)$, 
then we have (see e.g., Molchanov(2008))
\bqn{\label{taurho}{ \tau (Q_{\CE {A}})= \int_0^1 \frac{t(1-t)}{\CE {A}(t)} d \CE {A}'(t), \quad 
		\rho_S(Q_{\CE {A}})= 12 \int_0^1 \frac{1}{(1+ \CE{A}(t))^2} \, dt -3.
	}}

	\GM{ To illustrate the results stated above, we compare by simulations the dependence properties of both $C$  and $Q$. To this end, }
	we simulate random samples from both copulas and compute the empirical dependence measures. \GM{Specifically}, we generate a random sample from $C$ in which  Step 1-Step 4 in Subsection \ref{subsec:Sim_Algo} are repeated $10'000$ times.  Also, we simulate $\Lambda $ from \textbf{Model B} and two cases of $Q$ namely,  a Gumbel copula with parameter $10$ and a Clayton copula with parameter $10$.  
	Table \ref{table: depMes} describes the simulated empirical Kendall's tau \CE{and  Spearman's rho} for the random samples generated from $C$ and $Q$. \\
	\CE{
		\begin{center}
			\begin{tabular}{|c|| c|c| c| c||c|c|c|c|}
				\hline
				& \multicolumn{4}{c||}{  $Q$: Gumbel copula with $\alpha=\CE{10}$ } & \multicolumn{4}{c|}{ $Q$: Clayton copula  with $\alpha=\CE{10}$  }\\
				\hline
				$\EE{\Lambda} $   &  $\tau(C)$ & $\tau(Q)$ &  $\rho_S(C)$ & $\rho_S(Q)$ &     $\tau(C)$ & $\tau(Q)$ & $\rho_S(C)$ & $\rho_S(Q)$ \\
				\hline
				10 &   0.9059 & 0.9022  & 0.9871  & 0.9862  & 0.3533 &0.8343   &0.5030  &0.9588 \\
				\hline
				100 &    0.8980   &  0.9002 & 0.9848  & 0.9854 &0.0518  &0.8348   &0.0775  &0.9589\\
				\hline
				1'000 & 0.9007   &0.9004  & 0.9856& 0.9856   &0.0043    &0.8334  &0.0064 &0.9577 \\
				\hline
				10'000 & 0.9016   &0.9018   & 0.9857& 0.9859 &0.0019   &0.8324  &  0.0027& 0.9573\\
				\hline
				100'000 & 0.8997      &0.8996   &0.9851 &0.9854  &-0.0104 & 0.8316 &-0.0156  &0.9569 \\
				\hline
			\end{tabular}
			\captionof{table}{Empirical Kendall's Tau and Spearman's rho according to $\EE{\Lambda}$.} \label{table: depMes}
		\end{center}
	}
	
	\COM{
		\begin{center}
			\begin{tabular}{|c|| c|c|| c|c||c|c|}
				\hline
				& \multicolumn{2}{c||}{  $Q$: Gumbel copula $\alpha=\CE{10}$ } & \multicolumn{2}{c||}{ $Q$: Frank copula  $\alpha=\CE{5}$  }  & \multicolumn{2}{c|}{ $Q$: Clayton copula  $\alpha=\CE{10}$  }\\
				\hline
				$\EE{\Lambda} $   &  $\tau(C)$ & $\tau(Q)$ &  $\tau(C)$ & $\tau(Q)$ &  $\tau(C)$ & $\tau(Q)$ \\
				\hline
				10 &   0.9037 & 0.9003  &  0.3186 &  0.6645 &  0.3466 & 0.8312 \\
				\hline
				100 &    0.9006   & 0.9000  & 0.0564 &  0.6656 &  0.0472 & 0.8335\\
				\hline
				1'000 &     0.9002 & 0.9006 & 0.0007   &  0.6658  & 0.0139 & 0.8320 \\
				\hline
				10'000 &  0.8990   & 0.9004  & -0.0007 &  0.6653 & -0.0094 &  0.8336 \\
				\hline
				100'000 &   0.8965     &  0.9002 & -0.0058 &  0.6649 &-0.0053  & 0.8360 \\
				\hline
			\end{tabular}
			\captionof{table}{Empirical Kendall's Tau according to $\EE{\Lambda}$. } \}
		\end{center}
	}
	\CE{\ebe{The table above shows that} for the Gumbel copula case}, the level of dependence of a bivariate risk  governed by $C$ is lower or approximately equal to the one corresponding to $Q$ \CE{when $\EE{\Lambda} $ increases}.  For the case of \CE{Clayton}  copula,  the bigger $\EE{\Lambda} $, the weaker the dependence associated with $C$. In particular, for a copula $Q$ \ehe{with no} upper tail dependence, Clayton copula in our example, it can be seen that when $\EE{\Lambda} $ increases, $C$ tends to the independence copula. However, when $Q$ is an extreme value copula, Gumbel copula in our illustration, the rate of decrease in the level of dependence with respect to  $\EE{\Lambda} $ is small. These empirical findings are due to the correlation order demonstrated in \eqref{eq: corr_Order}. 
	\GM{To verify the results} \GM{ obtained from simulations, } \GM{ we show that, under (\ref{taurho}), for $\alpha=10$}, we obtain $\tau(\CE{Q_A })=0.9$ and $\rho_S(\CE{Q_A })=0.9855$ \GM{for the Gumbel copula} which are in line with the simulation results \ece{presented} in Table \ref{table: depMes}.\\
	\GM{It should be noted that for the Gumbel copula, the Pickands \ehe{dependence} function can be written as follows 
		\CE{
			$$ A(t)= (t^{1/\alpha}+ (1- t)^{1/\alpha})^{\alpha}, \quad t\in (0,1), \alpha \in (0,1) $$
		}
		leading to a closed form for $\tau (\CE{Q_A })$ \ebe{given by}
		$$\tau (\CE{Q_A })=  1-\frac{1}{\alpha}.$$  
	}
	Also, it is well-known that for \CE{Clayton} copula \eqref{copA} holds with \CE{$Q_A$} being the \ehe{independence}  copula, 
	hence for this case by \eqref{tau} we have $\limit{n}\tau(C_n)=0$, which confirms the findings in Table 1.

This section is concerned with the extremal properties of the df $F$ introduced in \eqref{intro} in terms of $G$ and $\Lambda$. 
The natural question which we want to answer here is \ehe{ whether the} extremal properties of $G$ and $F$ are the same. 
Therefore, we shall assume that $G$ is in the max-domain of attraction of some max-stable bivariate distribution $H$. Without loss of generality we shall assume that $H$ has unit \FRE  marginal df's. Hence, our assumption is that 
\BQN\label{md1}
 \limit{n} G^n (n x,ny)=H(x,y), \quad x,y\in [0,\IF).
\EQN
The max-stability of $H$ and the fact that its  marginal df's are unit \FRE  imply 
\BQN\label{md2} 
H^t(\ehe{t}x,\ehe{t}y)= H(x,y), \quad \forall x,y,t \in (0,\IF)
\EQN
see e.g., Falk(2010).
In case $\Lambda$ is a shifted geometric random variable as in {\bf Model A}, then the above assumptions imply 
for any $x,y$ non-negative (set $q:=1- \theta$) 
\BQNY
n[1- F(  n x,ny)]&=& n  \Bigl[ 1- \frac{ \theta G(nx,ny)}{1- qG(nx,ny)} \Bigr]\\
&=& n \frac{ 1- G(nx,ny)}{1- qG(nx,ny)} \\
&\to&   - \frac 1 \theta \ln H(x,y),  \quad n\to \IF.
\EQNY
Hence 
\BQN
 \limit{n}   F^n (  n x, ny )=  H^{1/\theta}(x,y)
\EQN
or equivalently, using \eqref{md2} 
\BQN
 \limit{n}  F^n ( nx/\theta, ny/\theta)= H^{1/\theta}(x/\theta, y/\theta)= H(x,y), \quad x,y\in (0,\IF)
\EQN
and thus $F$ is also in the same max-domain of attraction as $G$.

Our result below shows that the extremal properties of $G$ are preserved for the general case when $\EE{\Lambda}$ is finite. 
This assumption is natural in collective models, since otherwise we cannot insure such portfolios.

\def\aA{\overline{\theta}}

\BS\label{prop1}
If $\EE{\Lambda}$ is finite then $\mu_Q= \mu_C$. Moreover, if \eqref{md1} holds, then 
\BQN\label{conv}
 \limit{n}  F^n ( a_n x, a_n y)&=&H(x,y), \quad x,y\in (0,\IF),
\EQN
where $a_n= \EE{\Lambda} n$. 
\ES

\BRM i) 
It is well-known, see e.g., Falk (2010), that if $G$ is in the max-domain of attraction of $H$, then the coefficient of 
upper tail dependence $\mu_Q$ of $G$ with copula $Q$ exists and 
\BQNY\label{lambdaQ}
 \mu_Q &=& 2 + \ln H(1,1) = 2- 2 A(1/2).
 \EQNY
By the above proposition, $F$ is also in the max-domain of attraction of $H$, and thus 
\BQN\label{LQ}
\mu_C&=& 2- 2A(1/2)= \mu_Q \in [0,1].
\EQN
ii) Although $F$ and $G$ are in the same max-domain of attraction, the above proposition shows that the normalising constant $a_n= \EE{\Lambda}n$ for $F$ is different that for $G$ (here $a_n=n$)  if $\EE{\Lambda} \not=1$. 
\ERM

\section{Parameter Estimation and  Monte Carlo Simulations}

\subsection{\ebe{Parameter} Estimation} 
This section focuses on the estimation of the parameters of the new copula $C$ i.e., $\theta$ of $N$  and $\alpha$ of the  copula $Q$. Hereafter, we denote $\Theta=(\theta,\alpha)$.
There are three widely used methods for the estimation of the copula parameters. The classical one is the maximum likelihood estimation  (MLE). Another popular method is the inference function for margins (IFM), \ebe{which is} a step-wise parametric method.  First, 
the parameters of the marginal df's are estimated and then  the copula parameter \ehe{$\Theta$} are obtained by maximizing the likelihood function of the copula with the marginal parameters replaced by their first-stage estimators. 
\ebe{Typically,} the success of this method depends upon finding appropriate parametric models for the marginals, see Kim et al. (2007).\\
Finally, the  pseudo-maximum likelihood (PML) method, introduced by Oakes (1989) \ebe{consists \ece{also} of two steps.} 
In the first step, \ebe{the marginal df's are estimated non-parametrically.} The copula parameters are determined in the second step by maximizing the pseudo log-likelihood function. Specifically, let $X  \sim \ebe{G_1} $ and $Y \sim \ebe{G_2} $ where $\ebe{G_1} $ and $\ebe{G_2} $ are the unknown marginals df's of $X$ and $Y$.  For instance, if the data is not censored, a commonly used non-parametric estimator of $\ebe{G_1} $ and $\ebe{G_2} $  is  their sample empirical distributions which are specified as follows
\BQN \label{eq:empiricaldf}
\ebe{\widehat{G_1}} (x) =\frac{1}{n} \sum_{i=1}^{n} \mathbf{1}(X_i \leq x), \quad
 \ebe{\widehat{G_2}} (y) =\frac{1}{n} \sum_{i=1}^{n} \mathbf{1}(Y_i \leq y).
 \EQN
Therefore, in order to estimate the parameter $\Theta$, we maximize the following pseudo log-likelihood function 
\BQN\label{eq:CopulaF}
 l(\Theta) = \sum_{i=1}^{n} \ln c_\Theta (U_{1i},U_{2i}), \quad U_{1i}=\frac{n}{n+1}\ebe{\widehat{G_1}} (x_i), \quad  U_{2i}=\frac{n}{n+1}\ebe{\widehat{G_2}} (y_i),
\EQN
where $c_\Theta$ denotes the pdf of the copula. This rescaling is used to avoid difficulties arising from the unboundedness of the pseudo log-likelihood \ehe{function} \green{in \eqref{eq:CopulaF}} as $\ebe{\widehat{G_1}} (x_i)$ or $\ebe{\widehat{G_2}} (y_i)$ tends to 1, see Genest et al. (1995).\\
Kim et al. (2007)  show in a recent simulation study that the PML approach is better than the well-known IFM and MLE methods when the marginal df's are unknown, which is almost always the case in practice. Moreover, it is shown  in Genest et al. (1995) that the resulting  estimators from the PML approach are consistent and asymptotically normally distributed. \\
Therefore, for our study, we shall use the PML method for the estimation of  $\Theta$ which takes into account the empirical counterparts of the marginal df's to find the parameter estimators.\\
As described in the Introduction, we consider three types of distributions for the random variable $\Lambda$:
\BIT
\item \textbf{Model A:} $\Lambda$ follows a shifted Geometric distribution \ebe{with parameter $\theta \in (0,1)$}.\\
The pdf of the Geometric copula is given by\\
\BQN \label{CopulaF0}
c_\Theta(u_1,u_2)
&=&
W(v_1,v_2)
\Biggl(
\frac{(1-(1-\theta) v_1)^2(1-(1-\theta) v_2)^2}{\theta(1-(1-\theta) Q_\alpha (v_1,v_2))^3}
\Biggr) ,
\EQN
where $v_i=\frac{u_i}{\theta + (1-\theta)\CE{u_i} }, i=1,2$ and 
$$
W(v_1,v_2)=
(1-(1-\theta) Q_\alpha(v_1,v_2))
 \Bigl(\frac{\partial^2 Q_\alpha(v_1,v_2)}{\partial v_1 \partial v_2}\Bigr) \Bigr)
   +2(1-\theta)   \Bigl(\frac{\partial Q_\alpha(v_1,v_2)}{\partial v_1} \frac{\partial Q_\alpha(v_1,v_2)}{\partial v_2} \Bigr),$$
which yields the following pseudo log-likelihood function
\BQN\label{CopulaF2}
l(\Theta)\notag
&=& \sum_{i=1}^{n}
 \Bigl(2\ln(1-(1-\theta)v_{1i})+2\ln(1-(1-\theta)v_{2i})- \ln(\theta)-3\ln(1-(1-\theta)Q_\alpha(v_{1i},v_{2i}))+\ln W(v_{1i},v_{2i}) \Bigr).
\EQN

\COM{
with $W_i=(1-(1-\theta)Q_\alpha(v_{1i},v_{2i}))\frac{\partial^2 Q_\alpha(v_{1i},v_{2i})}{\partial v_{1i} \partial v_{2i}}+2(1-\theta)\frac{\partial Q_\alpha(v_{1i},v_{2i})}{\partial v_{1i}}\frac{\partial Q_\alpha(v_{1i},v_{2i})}{\partial v_{2i}}$.\\
}
\item  \textbf{Model B:} $\Lambda$ follows a Shifted Poisson distribution with parameter $\theta>0$.\\
The pdf of the shifted Poisson copula is of the form

\BQN \label{eq:ShiftedPoissonpdf}
c_\Theta(u_1,u_2)
&=&
W(v_1,v_2)
\Biggl(\frac{e^{\theta (Q_\alpha(v_{1},v_{2})+1-v_{1}-v_{2})}}{(1+\theta v_1)(1+\theta v_2)} \Biggr)  ,
\EQN
where $v_j=f^{-1}(u_j)$ with $f(x)= x \exp( \theta( x-1) )$ and 
$$W(v_1,v_2)=
(1+\theta Q_\alpha(v_1,v_2))
\Bigl( \frac{\partial ^2 Q_\alpha(v_1,v_2)}{\partial v_1 \partial v_2}\Bigr)
+\theta(2+\theta Q_\alpha(v_1,v_2)) \Bigl(  \frac{\partial Q_\alpha(v_1,v_2)}{\partial v_1} \frac{\partial Q_\alpha(v_1,v_2)}{\partial v_2} \Bigr).$$
 
The corresponding pseudo log-likelihood of the above copula is thus given by

\BQNY
l(\Theta)
&=& \sum_{i=1}^{n} \Bigl(\theta (Q_\alpha(v_{1i},v_{2i})+1-v_{1i}-v_{2i})-\ln (1+\theta v_{1i}) -\ln (1+\theta v_{2i})+\ln W(v_{1i},v_{2i}) \Bigr).
\EQNY

\COM{
with $W(v_1,v_2) =(1+\theta Q_\alpha(v_{1i},v_{2i}))
\frac{\partial Q_\alpha(v_{1i},v_{2i})}
{\partial v_{1i} \partial v_{2i}}
+(2+\theta Q_\alpha(v_{1i},v_{2i}))
\theta \frac{\partial Q_\alpha(v_{1i},v_{2i})}{\partial v_{1i}} \frac{\partial Q_\alpha(v_{1i},v_{2i})}{\partial v_{2i}} $\\
}

\item  \textbf{Model C:}  $\Lambda$ follows a Truncated Poisson distribution with parameter $\theta>0$.\\
The joint  density of the truncated Poisson copula is given by
\BQN\label{eq: TruncatedPoisPdf}
c_\Theta(u_1,u_2)
&=& \frac{1 }{\theta} (1- e^{-\theta}) W(v_{1},v_{2}) e^{\theta(1-v_1-v_2 + Q_\alpha(v_1,v_2)) } ,
\EQN

where 
$$W(v_1,v_2) = \theta \frac{\partial Q_\alpha(v_1,v_2)}{\partial v_1} \frac{\partial Q_\alpha(v_1,v_2)}{\partial v_2}+\frac{\partial^2 Q_\alpha(v_1,v_2)}{\partial v_1 \partial v_2}, $$ 
$$v_j= \frac{1}{\theta} \ln \biggl(  1 + \frac{u_j(1-e^{-\theta}) }{e^{-\theta}}\biggr), \quad j=1,2. $$
The resulting pseudo log-likelihood of the above copula can be written as follows

\BQN\label{eq:TruncatedPoissonpdf}\notag
l(\Theta)
&=& \sum_{i=1}^{n}\Bigl( \ln\Bigl(\frac{1-e^{-\theta}}{\theta} \Bigr)+ \theta(1-v_{1i}-v_{2i}) + \theta Q_\alpha(v_{1i},v_{2i})+ \ln W(v_{1i},v_{2i})\Bigr).
\EQN

\EIT
\COM{
with $ W=\theta \frac{\partial Q_\alpha(v_{1i},v_{2i})}{\partial v_{1i}} \frac{\partial Q_\alpha(v_{1i},v_{2i})}{\partial v_{2i}}+\frac{\partial^2 Q_\alpha(v_{1i},v_{2i})}{\partial v_{1i} \partial v_{2i}}$.

To obtain the estimator $\hat{\theta}$ of $\theta$, we need to maximize the pseudo log-likelihood of the copula density and we get:\\
\BQN
\hat{\theta}= \underset{\theta} {\text{argmax}} L(\theta)
\EQN 
}
\GM{
\BRM \label{rem:copulaConv}
		The copula $C_{\theta}$ of Model A and Model B include the corresponding original copula $Q$. In particular, if $\theta=1$ the pdf $c_{\theta}$ in \eqref{CopulaF0}  becomes the pdf  of the original copula $Q$, see e.g., Zhang and  Lin (2016), while the copula  $C_{\theta}$ of Model B  reduces to the original copula $Q$ when $\theta=0$.
\ERM
}
Next, we generate  random samples from the proposed copula models $C$. 
\subsection{Monte Carlo Simulations} \label{subsec:Sim_Algo}
Based on the distributional properties of $F$ derived in Section 2, we have the following pseudo-algorithm for 
the simulation procedure which depends on the choice of  $\Lambda$ and $Q$:
\BIT
	\item \textbf{Step 1:} Generate a \ebe{value} $\lambda$  from  $\Lambda$.
	 \item \textbf{Step 2:} Generate   $\lambda$ random samples $(U_{1,i},U_{2,i}),i=1,\ldots,\lambda,$  from the original copula $Q$. 
	  \item \textbf{Step 3:}  Calculate  $ (M_{1},M_{2} )$ as follows 
	  $$ M_{j}= \max_{i=1,\ldots,\lambda} U_{j,i}, \quad j=1,2.$$ 
	   \item \textbf{Step 4:} Return $ (V_{1},V_{2} )$, such that  
	   $$ V_j= L_\Lambda(- \ln M_j ),  \quad  j=1,2. $$
\EIT 
Simulation results are important for exploring the dependence of $F$. The simulation results in the 
table below complete those presented already in Table \ref{table: depMes}. 
\CE{  \green{In this regard}, we generate random samples from the Joe copula with parameter $\alpha=10$.}
\CE{\begin{center}
	\begin{tabular}{|c|| c|c| c| c|}
	\hline
	& \multicolumn{4}{c|}{  $Q$: Joe copula with $\alpha=10$ } \\
\hline
$\EE{\Lambda} $   &  $\tau(C)$ & $\tau(Q)$ &  $\rho_S(C)$ & $\rho_S(Q)$  \\
\hline
10 & 0.8982  & 0.8194  &0.9849  & 0.9504   \\
\hline
100 & 0.9005   &0.8190  &0.9857   &0.9509  \\
\hline
1'000 &0.8997    &0.8164  & 0.9855 & 0.9492  \\
\hline
10'000 &0.9004    & 0.8209  &0.9857  & 0.9520 \\
\hline
100'000 &  0.8999     &0.8206   &0.9852  & 0.9513\\
\hline
		\end{tabular}
	\captionof{table}{Empirical Kendall's Tau and Spearman's rho according to $\EE{\Lambda}$. } \label{table: depMes2}
\end{center}}

For the Joe copula, the Pickands dependence function can be written as follows
\BQN\notag
A(t)= 1-((\psi_1(1-t))^{-\alpha}+(\psi_2 t)^{-\alpha})^{-\frac{1}{\alpha}}
\EQN
where $\psi_1, \psi_2 \leq 1$  , $t \in (0,1)$ and $\alpha \in (0,1)$. \\
By using \eqref{taurho} and for $\alpha=10$ and $\psi_1=\psi_2=1$,  we obtain \ehe{$\tau(Q_{A})=0.9066$} and $\rho_{S}(Q_{A})=0.9874$ which \ehe{are in line with the simulation results observed  in Table \ref{table: depMes2}  for $\tau (C)$ and $\rho_S(C)$ as $\EE{\Lambda} $ increases.}\\

Another benefit of our simulation algorithm is that we can assess the accuracy of our estimation method proposed above. 
\green{
Therefore, we simulate random samples of size $n$ from the copula $C$ with different distributions  for $\Lambda$: \textbf{Model A}, \textbf{Model B} and \textbf{Model C} and two types of copula for $Q$: the Gumbel copula and the Joe copula.   Hereof, the parameters  $\theta$ of $\Lambda$ and $\alpha$ of $Q$   are estimated from the dataset described in Subsection \ref{subsec:data_VG} and are presented in Table \ref{table: Parameters} .  }
\green{
\begin{center}
	\begin{tabular}{|c|| c|c|| c| c|}
	\hline
	& \multicolumn{2}{c||}{  $Q$: Joe copula  } & \multicolumn{2}{c|}{  $Q$: Gumbel copula  } \\
\hline
 Model for $\Lambda$   & $ \theta$  & $\alpha$ & $ \theta$  & $\alpha$  \\
\hline
Model A &  0.3254 & 2.3727  &0.7630  & 2.2758   \\
\hline
Model B & 0.9537  & 2.6634  &0.1490  & 2.3276  \\
\hline
Model C &  1.8660   & 2.5885  & 0.3133 & 2.3240  \\
\hline
		\end{tabular}
	\captionof{table}{Parameters used for sampling from $C$. } \label{table: Parameters}
\end{center}
}

\green{
\begin{center}
	\begin{tabular}{|c|| c|c| c|c || c| c | c|c || c| c | c|c |}
	\hline
	& \multicolumn{4}{c||}{  Model A  } & \multicolumn{4}{c|}{  Model B } & \multicolumn{4}{c|}{  Model C }\\
\hline
$n$  & $ \hat{\theta}$  &  Diff.  & $ \hat{\alpha}$  & Diff.   & $ \hat{\theta}$  &  Diff.  & $ \hat{\alpha}$  & Diff.  & $ \hat{\theta}$  &  Diff.  & $ \hat{\alpha}$  & Diff.  \\
\hline
100 & 0.2461  &-24$\%$    &2.2255  &-6$\%$   &1.0765  &13$\%$ &2.6597&0$\%$ &1.7400 &-7$\%$ &2.2535 & -13$\%$  \\
\hline
1'000 &0.3353   &3$\%$    &2.3262  & -2$\%$  &0.9906  &4$\%$ &2.6999&1$\%$ &1.9238 & 3$\%$ &2.6491 &   2$\%$\\
\hline
10'000 &0.3304   &2$\%$    & 2.3260 &-2$\%$   & 0.9795 &3$\%$ &2.6651&0$\%$ &1.8996 &2$\%$ &2.5999 & 0$\%$  \\
\hline
100'000 &0.3285   &1$\%$    &2.3462  & -1$\%$  & 0.9541 &0$\%$ &2.6600& 0$\%$ &1.8721 & 0$\%$ &2.5877 & 0$\%$   \\
\hline
		\end{tabular}
	\captionof{table}{Parameters used for sampling from $C$ where $Q$ is the Joe Copula. } \label{table: est2}
\end{center}
}
\green{
\begin{center}
	\begin{tabular}{|c|| c|c| c|c || c| c | c|c || c| c | c|c |}
	\hline
	& \multicolumn{4}{c||}{  Model A  } & \multicolumn{4}{c|}{  Model B } & \multicolumn{4}{c|}{  Model C }\\
\hline
$n$  & $ \hat{\theta}$  &  Diff.& $ \hat{\alpha}$  & Diff.   & $ \hat{\theta}$  &  Diff. & $ \hat{\alpha}$  & Diff.  & $ \hat{\theta}$  &  Diff.  & $ \hat{\alpha}$  & Diff.  \\
\hline
100 &0.9565   &25$\%$   &2.3595  &4$\%$  &0.1712  &15$\%$ &2.4308&4$\%$ &0.3164 & 1$\%$ &2.3675 & 2$\%$  \\
\hline
1'000 &0.7376   &-3$\%$   &2.3076  &1$\%$  &0.1563 & 5$\%$&2.3458 &1$\%$ &0.3084 & -2$\%$&2.3126 &0$\%$  \\
\hline
10'000 &0.7660   &0$\%$    &2.3083  &1$\%$  & 0.1545  &4$\%$ &2.3476& 1$\%$&0.3136 & 0$\%$&2.3185 &0$\%$  \\
\hline
100'000 &0.7596  &0$\%$    &2.2639  &-1$\%$   & 0.1506  &1$\%$ &2.3232& 0$\%$ &0.3279 &5$\%$ &2.3063 &-1$\%$   \\
\hline
		\end{tabular}
	\captionof{table}{Parameters used for sampling from $C$ where $Q$ is the Gumbel Copula. } \label{table: est1}
\end{center}
}
\green{
It can be seen from Table \ref{table: est2} and  Table \ref{table: est1} that  the estimated parameters from the simulated samples tend to the true value of the parameters as the sample
size $n$ increases, thus indicating the accuracy of our proposed models.
}
\COM{{\bf Gaussian copula, FGM copula??}

\CE{ In this subsection, we simulate random samples from two copulae. The same procedure of Section 2 is followed where the above steps are repeated 10'000 times. The table below shows a simulation of random samples from Student copula with parameter $\alpha=0.5$ and $m=10$ degrees of freedom  and Gaussian copula with parameter $\alpha=0.5$.
\begin{center}
	\begin{tabular}{|c|| c|c| c| c||c|c|c|c|}
	\hline
	& \multicolumn{4}{c||}{  $Q$: Student copula with $\alpha=0.5$, $m=10$ } & \multicolumn{4}{c|}{ $Q$: Gaussian copula  with $\alpha=0.5$}\\
\hline
$\EE{\Lambda} $   &  $\tau(C)$ & $\tau(Q)$ &  $\rho_S(C)$ & $\rho_S(Q)$ &     $\tau(C)$ & $\tau(Q)$ & $\rho_S(C)$ & $\rho_S(Q)$ \\
\hline
10 &   0.2306 & 0.3375  & 0.3370  & 0.4822  & 0.2103 &0.3300   &0.3102 &0.4778 \\
\hline
100 &    0.1433   & 0.3464 & 0.2124  & 0.4941 &0.0845  &0.3282   &0.1264  &0.4747\\
\hline
1'000 & 0.1052   &0.3263  & 0.1563 & 0.4678  &0.0340    &0.3340  &0.0509 &0.4829 \\
\hline
10'000 & 0.1006   &0.3305   & 0.1498 & 0.4717 &0.0125   &0.3297  &  0.0186 & 0.4777\\
\hline
100'000 & 0.0717      &0.3359   &0.1072 &0.4815  &0.0109 &0.3277  &0.0162  &0.4745\\
\hline
		\end{tabular}
	\captionof{table}{Empirical Kendall's Tau and Spearman's rho according to $\EE{\Lambda}$. } \label{table: depMes2}
\end{center}
}

\CE{For verification purpose, the Pickands dependence function of the student copula is given by
\BQN\notag
\widetilde{A(s)}= s  t_{m+1} (k_s) +(1-s) t_{m+1} (k_{1-s})
\EQN 
where t is the pdf of a student distribution with parameters $(\alpha,m)$
and
\BQN\notag
k_s=\sqrt{\frac{1+m}{1-\alpha^2}}\Bigl(\Bigl(\frac{s}{1-s}\Bigr)^{\frac{1}{m}}-\alpha \bigr)
\EQN
}
\CE{leading to a closed form of the Kendall's $\tau$ dependence function $\tau(Q_{\widetilde{A}})= \frac{2}{\pi} arcsin(\alpha)$ and for the Spearman rank correlation $\rho_S$,                                             $\rho_S(Q_{\widetilde{A}})= \frac{6}{\pi} arcsin(\frac{\alpha}{2})$ \\}
\CE{By using \eqref{taurho} and for $\alpha=0.5$ and $m=10$,  we get $\tau(Q_{\widetilde{A}})=0.3333$ and $\rho_{S}(Q_{\widetilde{A}})=0.4826$ }
}
\GM{
		\subsection{Influence of  $ X_{N:N} + Y_{N:N} $ on total loss } 
		In this subsection, we focus on  the distribution of the aggregate claim of  two insurance portfolios by excluding the largest claim of each  portfolio. Specifically, we analyse the aggregate influence of  $ M_N:=X_{N:N} + Y_{N:N} $   on  some risk measures of the total loss  
		$ S_N = \sum_{i=1}^N (X_i + Y_i) $.   Moreover,  by considering the joint distribution of $(X_{N:N}, Y_{N:N})$ we quantify the individual impact of $X_{N:N}$  and $Y_{N:N}$ on  the distribution of $ S_N$. Let $ S_N^*  $ be the aggregate claim excluding the largest claims, based on some risk measure $\rho(.)$ and \ehm{suppose that $X_i,Y_i$'s have a finite second moment,} the influence of the largest claims on the aggregate claim is evaluated as follows
		\BQNY
					\ehm{I^*}	= \rho(S_N) -  \rho(S_N^*).
		\EQNY
By the covariance capital allocation principle, the contribution of $X_{N:N}$ on the change of  the distribution of $S_N$ is given by 
		\BQNY
					I(X_{N:N},\MG{M_N}) = \frac{cov( X_{N:N}, M_N)   }{var( M_N)} \ehm{I^*}.
		\EQNY
To illustrate our results we have implemented the following simulation  pseudo-algorithm:
\BIT
	\item \textbf{Step 1:} Generate the number of claims $N$  from  $\Lambda$.
	 \item \textbf{Step 2:} Generate  $N$ random samples $(u_{1,i},u_{2,i}), i= 1,\ldots, N,$  from the original copula $Q$.
	 \item \textbf{Step 3:}  For each portfolio, simulate $N$  claim sizes by  using the inverse method as follows
	 						$$ X_i^{b}= F_1^{-1}(u_{1,i}), \quad  Y_i^b= F_2^{-1}(u_{2,i}) , i= 1,\ldots, N,$$
	 						where $ F_i, i=1,2,$ is the df of $X$ and $Y$, respectively. 
	  \item \textbf{Step 4:}	Evaluate the total loss with and without the largest claims, respectively 		
	  $$ 	S_N^b= \sum_{i=1}^N (X_i^{b} + Y_i^{b}), \quad S_N^{*b}. $$		
\EIT
To obtain the simulated distribution of $S_N$ and $S_N^{*}$ \textbf{Step 1-4}  are repeated $B$ times . 
The results presented in Table \ref{table: largestClaims} is in million and is obtained from the following assumptions: 
\BIT
	\item  number of simulations $B= 100'000$, 
	\item  the original copula is a Gumbel copula with dependence parameter $\alpha=2.324$,
	\item  the number of claims follows the Shifted Poisson (\textbf{Model B})  with parameter $\theta=1000 ,$
	\item  the claim sizes  are Pareto distributed as follows
	$$X_i \sim Pareto(10000, \MG{2.2} ), \quad  Y_i \sim Pareto(50000, \MG{2.5} ). $$
\EIT
\MG{
\begin{center}
	\begin{tabular}{|c| c|c| c| c|c|c|}
	\hline
Risk measures  &   $S_N$ & $S_N^*$ &  $I^*$  & $I^*$ (in $\%$) &  $ I(X_{N:N},\MG{M_N}) $ & $ I(Y_{N:N},\MG{M_N}) $  \\
\hline
 Mean   & 101.77 & 100.21 & 1.57 & 1.54 & 0.38 & 1.19   \\
 \hline
 Standard deviation   &    4.41 & 3.91 &  0.50   & 11.24 & 0.15 &  0.35  \\
  \hline
 VaR (99 \%)   & 112.75  & 109.65 & 3.10 &  2.74 & 0.90 & 2.20 \\
   \hline
 TVaR (99 \%)   &  117.08   & 111.03 & 6.05 & 5.17 &1.75& 4.30\\
    \hline
		\end{tabular}
	\captionof{table}{Influence of the largest claims on the total loss.  } \label{table: largestClaims}
\end{center}
}
It can be seen that a significant proportion of the aggregate claims is consumed by $ X_{N:N} + Y_{N:N} $. For instance, based on the standard deviation as risk measure, \MG{$11.24 \%$}  of the total loss is driven by the largest claims. In this regards,  $ X_{N:N}$ has more important contribution  to  $I^*$ than $ Y_{N:N}$.  This result is helpful for the insurance company when  choosing the appropriate reinsurance treaty in the sense that the main source of volatility of the correlated portfolios is quantified. 
		}	
\section{Real insurance data applications}
In this section, we illustrate the applications of the new copula families \ehe{in the modelling of} 
\green{three}  real insurance  data. \ehe{Specifically, we shall} consider four  copula families for $Q_\alpha$: Gumbel, Frank, Student and Joe and three mixture copulas in which $\Lambda$  with parameter $\theta$ follows one of the three distributions: \ehe{Shifted} Geometric, Shifted Poisson and Truncated Poisson. 
The AIC criteria is used to assess the quality of each model \ehe{fit} relative to each of the other models.
\COM{
firstly on the workers compensation insurance data which have been used in  Zhang et al. (2015) and secondly  on  two types of  insurance loss and the corresponding allocated loss adjustment expenses (ALAE). The first loss ALAE dataset is retrieved from an insurance company and the second one is from the insurance Services Office which is available on R.

 the  that is implemented in the R package. 
The mixture copula that best fits the data is the Gumbel Shifted Poisson copula as its AIC is the smallest followed by the Gumbel Truncated Poisson. The below table presents the estimation results and AIC values for the different copulas.
}

\subsection{Loss ALAE from accident insurance} \label{subsec:data_VG}

\ehe{ We shall model} real insurance data  from a large insurance company operating in Switzerland.  The dataset consists of  \ehe{33'258} accident insurance losses and their corresponding allocated loss adjustment expenses (ALAE) which includes mainly the cost of medical consultancy and legal fees. The observation period  encompasses  the claims occuring  during the accident period  1986-2014\footnote{Data set can be downloaded here 
\url{http://dx.doi.org/10.13140/RG.2.1.1830.2481}}.\\

Let $X_i$ be the $i^{th}$ loss observed and $Y_i$ its corresponding ALAE. \\
Some statistics on the data are summarised in Table \ref{table: Vaudoise Stat}. \\
\begin{center}
	\begin{tabular}{|c|c|c|}
		\hline
& Loss & ALAE \\

\hline
 \hline
Min  &10  & 1  \\
\hline 
Q1  &  13'637  & 263      \\
\hline 
Q2  & 32'477     &  563       \\
\hline 
Q3  & 95'880      & 1'509       \\

\hline 
Max  & 133'578'900     & 2'733'282        \\
\hline
\hline
No. Obs. & 33'258   &  33'258   \\ 
\hline
Mean &   292'715   &    5'990 \\ 
\hline
Std. Dev. & 2'188'622      &  42'186    \\ 
\hline
		\end{tabular}
	\captionof{table}{Statistics for Loss ALAE data from accident line. } \label{table: Vaudoise Stat}
\end{center}

\begin{figure} [H]
	\centering
	\includegraphics [scale=0.35]{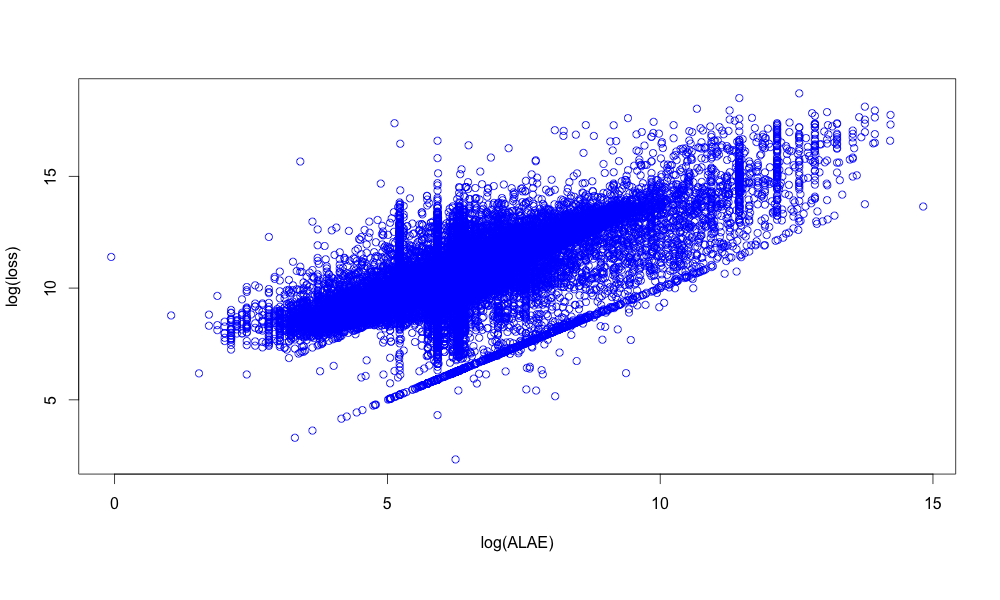}
	\caption{ Scatterplot for log ALAE and log Loss: accident insurance data. }
		\label{Figure: ScatterPlot_Xi_Yi}
	\end{figure}

The scatterplot of (ALAE, loss) on a log scale is depicted in Figure   \ref{Figure: ScatterPlot_Xi_Yi}. It can be seen that   large values of loss is likely to be associated with large values of ALAE. In addition, the empirical estimator of some dependence measures in \ehe{Table \ref{table:dep_Mesures_Vaudoise}} suggests a positive dependence between $X_i$ and $Y_i$. For instance, the empirical estimator of the upper tail dependence of 0.6869 indicates that there is a strong dependence in the tail of the distribution of $X_i$ and $Y_i$.  \\

\begin{center}
	\begin{tabular}{|c|c|}
  \hline
Pearson's Correlation  &  0.7460      \\
  \hline
Spearman's Rho  &  0.7465  \\
  \hline
Kendall's  Tau &  0.6012  \\
\hline
Upper tail dependence  &  0.6869   \\
\hline
 	\end{tabular}
	\captionof{table}{Empirical dependence measures for  Loss ALAE data from accident line. }\label{table:dep_Mesures_Vaudoise}
\end{center}

Referring to the marginal's estimator in \eqref{eq:empiricaldf},  the estimation results for each copula model are found by maximizing \eqref{eq:CopulaF} and are summarized in Table \ref{table:parm_Vaudoise} below. \\
\begin{center}
	\begin{tabular}{|l|c|c|c|c|}
		\hline
		 Model & $\theta$  & $\alpha$ & $m$ &   AIC   \\

		\hline
		Gumbel             &   -     &  2.3876   & - &  -32'073    \\
		
		Gumbel Geometric       &   0.7630  &  2.2758   & - &   -32'128   \\
		
		Gumbel Truncated Poisson  &  0.3133   &  2.3240    & - &  -32'104   \\
		
		Gumbel Shifted Poisson  & 0.1490   & 2.3276  & - & -32'059   \\
		\hline
                      Frank                  &  -    &  8.0774& -  &   -30'137   \\
                    
                      Frank Geometric                 &  0.9999  &  8.0772  &- &  -30'134   \\
                     
                      Frank Truncated Poisson       &  0.0001   &  8.0773  &-&  -30'135   \\
                     
                      Frank Shifted Poisson             &    0.0001  &  8.0773  &- &  -30'135   \\
                     \hline
                     Student                & -  & 0.8142   & 1.9805 &   -32'909  \\
                     
                     Student Geometric       &  0.1137&  0.5492& 1.9992 &-38'088   \\        
                    
                    Student Truncated Poisson       & 0.0001 & 0.7841 & 9.6744 & -28'672   \\       
                     
                    Student Shifted Poisson               & 0.0001 & 0.7885 & 8.7113 & -29'042   \\
                     \hline
                     Joe                  &    -  & 3.0967 & - & -30'655 \\ 
                     
                     Joe Geometric                & 0.3254   & 2.3727  &- & -33'015 \\
                     
                     Joe Truncated Poisson                 & 1.8660  & 2.5885 & - & -32'578 \\
                     
                     Joe Shifted Poisson                 & 0.9537  & 2.6634  & - & -32'411 \\
                     \hline
		\end{tabular}
	\captionof{table}{Copula families parameters estimates.} \label{table:parm_Vaudoise}
\end{center}
It can be seen that  the model which best fits the data is the Student Geometric copula followed by the Joe Geometric copula. We note in passing that the Student copula $Q_\alpha$  has an additional parameter $m$ which is the degree of freedom.
 
\subsection{Loss ALAE from general liability insurance}
This data set describes the general liability claims associated with their ALAE retrieved from the Insurance Services Office available in the  R package.
In this respect, the sample consists of  1'466 uncensored data points and 34 censored observations.   We refer to Denuit et al. (2006)  for more details on the description of the data. 
Let $X_i$ be the $i^{th}$ loss observed and $Y_i$ the ALAE associated to the settlement of $X_i$.
Each loss is associated with a maximum insured claim amount (policy limit) $M$. 
Thus, the loss variable $X_i$  is  censored when it exceeds the policy limit  $M$. We define the censored indicator of the loss variable by
	$$\delta_i = 
	 \left\{
			 	   \begin{array}{lcl}
         			 1 & \mbox{if} & X_i \leqslant M, \\
         		    0  & \mbox{if} &  X_i > M, i=1,\ldots,1'500.
              	\end{array}
      \right.
      $$
\ehe{Next, we shall use}  the  Kaplan-Meir estimator $\hat{G}_X$ to estimate  $\ebe{G_1} $  and the empirical distribution $\hat{G}_Y$ for $\ebe{G_2}$  as in \eqref{eq:empiricaldf}.  
In particular, the corresponding pseudo log-likelihood function is given by
\BQN\label{eq:censoredLikeli}
 l(\Theta) = \sum_{i=1}^{n}  
 \Bigl(
  \delta_i \ln (c_\Theta (U_{1i},U_{2i}) + (1-\delta_i) 
  \ln\Bigl(  1- \frac{C_\Theta (U_{1i},U_{2i})}{\partial U_{2i}} \Bigr)   
 \Bigr),
\EQN
where 
$U_{1i}= \frac{n}{n+1} \hat{G}_X (x_i)$ and $U_{2i}= \frac{n}{n+1} \hat{G}_Y (y_i) $ for $ i=1,\ldots,n$, see Denuit et al.(2006).
\COM{
Table \ref{table:dep_loss}  below summarizes the empirical dependence measures between $X_i$ and $Y_i$.
 \begin{center}
	\begin{tabular}{|c|c|}
  \hline
Pearson's Correlation  & 0.4519  \\
  \hline
Spearman's Rho  &0.4519  \\
  \hline
Kendall's  Tau &0.3154	  \\
\hline
Upper tail dependence  & 0.3777 \\
\hline
 	\end{tabular}
\captionof{table}{Dependence measures for  Loss ALAE data from general liability inusrance. }\label{table:dep_loss}
\end{center}

}
By maximizing \eqref{eq:censoredLikeli}, the resulting estimators of $\Theta$  for the considered copula models are presented in Table \ref{table:parm_Denuit}. \\
\begin{center}
	\begin{tabular}{|l|c|c|c|c|}
		\hline
		 Model & $\theta$  & $\alpha$ & $m$ &   AIC   \\

		\hline
		Gumbel             &      -  & 1.4284  & - & -210.18  \\
		
		Gumbel Geometric       & 0.5425   & 1.3127 & - & -278.23  \\
		
		Gumbel Truncated Poisson  & 0.0001 &  1.4422 & -  & -360.49  \\
		
		Gumbel Shifted Poisson  &  0.1410  & 1.4083  &-  &  -361.20  \\
		\hline
                      Frank                  &  -     & 3.0440 & - & -321.44 \\
                    
                      Frank Geometric                 & 0.7800  & 2.7464  &- & -174.40 \\
                     
                      Frank Truncated Poisson                 & 0.0001  & 3.0375  &- & -306.40\\
                     
                      Frank Shifted Poisson                 & 0.0001  & 3.0375  &- & -306.41 \\
                     \hline
                     Student                & -  & 0.4642  &10.0006 & -180.99 \\
                     
                     Student Geometric       &0.7095&0.4252 &9.1897&-228.82\\        
                    
                    Student Truncated Poisson       & 1 & 0.4094 &13.9922 & -271.40   \\       
                     
                    Student Shifted Poisson               &1 &0.4016 &13.9983 & -295.42  \\
                     \hline
                     Joe                  &  -     & 1.6183 &- & -179.00 \\
                     
                     Joe Geometric                 & 0.4379 & 1.3864  &-  & -292.41 \\
                     
                    Joe Truncated Poisson                 & 0.0607  & 1.6356  & - & -331.21 \\
                     
                    Joe Shifted Poisson                 & 0.8075  & 1.4629  & - & -361.76 \\
                     \hline
		\end{tabular}
	\captionof{table}{Copula families parameters estimates.} \label{table:parm_Denuit}
\end{center}
Since the Joe Shifted Poisson copula has the the smallest AIC it represents the best model for describing the dependence in the dataset followed by the Gumbel Shifted Poisson copula. 
\COM{
the Gumbel shifted Poisson copula outperforrms the Gumbel Geometric copula introduced by Zhang et al. (2015) as well as the other copulas defined above as its AIC measure is the smallest and thus represents a better model fit for the data.\\
}
\COM{
\subsection{Loss ALAE from medical  insurance}
In this section, we will be working on the SOA  Medical Group Insurance datasets\footnote{Datasets are available  on the Society of Actuaries website at \url{https://www.soa.org/Research/Experience-Study/group-health/91-92-group-medical-claims.aspx}} describing the medical claims observed over the years $1991-1992$. The $171'236$ claims recorded in both datasets are part of a larger database that includes the losses of $26$ insurers over the period 1991-1992, see \cite{Soa_Data} for more description on the data. \\
The two components which are of interest to us, are the hospital charges , corresponding to the loss variable $X_i$,  and the \green{other} charges corresponding to the ALAE variable $Y_i$ associated to the settlement of $X_i$. \\
The same dataset has been explored in \cite{cebrian2003analysis} where claims occurring during the accident year $1991$ have been considered. For our study, we are going to work with claims occurring during the accident year $1992$. The sample comprises of $75'789$ claims. There are $4$ different medical group plan types.  Each policyholder belongs to one of these medical groups. In this respect, we consider the \green{accident year}  $1992$  records relating to Plan type $4$, the loss variable $X_i$ exceeding $25'000$ in order to have a positive dependence between the loss and the ALAE, and strictly positive ALAE. Some statistics on the data are presented in Table \ref{table:stat_Health}. \\

\begin{center}
	\begin{tabular}{|c|c|c|}
		\hline
& Loss & ALAE \\
\hline
 \hline
Min  &25'003  & 5  \\
\hline 
Q1  &   30'859 &  7'775     \\
\hline 
Q2  & 40'985     &  14'111       \\
\hline 
Q3  & 64'067     & 23'547     \\
\hline 
Max  & 1'404'432     & 409'586       \\
\hline
\hline
No. Obs. & 5'106   &  5'106  \\ 
\hline
Mean &   62'589   &    20'001 \\ 
\hline
Std. Dev. &   69'539  & 24'130     \\ 
\hline
		\end{tabular}
	\captionof{table}{Statistics for Loss ALAE data from medical insurance. } \label{table:stat_Health}
\end{center}
The scatterplot of (ALAE, loss) on a log scale is depicted in Figure   \ref{Figure: ScatterPlot_Xi_Yi2}.  

\begin{figure} [H]
	\centering
	\includegraphics [scale=0.5]{Figure2.png}
	\caption{ Scatterplot for log ALAE and log Loss: medical  insurance data. }
		\label{Figure: ScatterPlot_Xi_Yi2}
	\end{figure}
Furthermore, the dependence measures in Table \ref{table:dep_measures_Med} indicates positive dependence between these two variables with an empirical upper tail dependence of $0.3805$. 
The empirical dependence measures between the losses and their respective ALAE  are shown in Table \ref{table:dep_measures_Med} below.\\
 \begin{center}
	\begin{tabular}{|c|c|}
  \hline
Pearson's Correlation  &  0.4442    \\
  \hline
Spearman's Rho  & 0.4442 \\
  \hline
Kendall's  Tau &	0.3088  \\
\hline
Upper tail dependence  & 0.3805  \\
\hline
 	\end{tabular}
	\captionof{table}{Dependence measures for  Loss ALAE data from medical insurance. }\label{table:dep_measures_Med}
\end{center}

By maximizing \eqref{eq:CopulaF},  we obtain the estimation results for the copula models summarized hereunder in Table \ref{table:parm_med}.\\
\begin{center}
	\begin{tabular}{|l|c|c|c|c|}
		\hline
		 Model & $\theta$  & $\alpha$ & $m$ &   AIC   \\

		\hline
		Gumbel             &  -  &1.4328   & - &-1'371.21   \\
		
		Gumbel Geometric       & 0.9751   &1.4262  & - &-1'369.28  \\
		
		Gumbel Truncated Poisson  &0.0380 & 1.4278  & -  &-1'369.27  \\
		
		Gumbel Shifted Poisson  &  23.7511 & 1.4035  &-  & -1'328.14  \\
		\hline
                      Frank                  &  -    & 3.0482  & - & -1'137.12 \\
                    
                      Frank Geometric                 & 0.9999 & 3.0481  &- & -1'135.09 \\
                     
                      Frank Truncated Poisson                 &0.0001   &3.0481  &- & -1'135.10 \\
                     
                      Frank Shifted Poisson                 &0.0001   &  3.0481 &- & -1'135.12 \\
                     \hline
                     Student                & -  &0.4576  &11.8933 & -1'195.83 \\
                     
                     Student Geometric       & 0.9999& 0.4576 & 11.8921& -1'193.82\\        
                    
                    Student Truncated Poisson       & 0.0001  & 11.8900  & 7.1017 &-1'193.83   \\       
                     
                    Student Shifted Poisson               &0.0001 & 0.4576 & 11.8902 & -1'193.82 \\
                     \hline
                     Joe                  &  -     & 1.6440  &- & -1'315.35 \\
                     
                     Joe Geometric                 &0.5804   & 1.4477   &-  &  -1'393.23 \\
                     
                    Joe Truncated Poisson                 & 1.0243  & 1.4832  & - & -1'386.87  \\
                     
                    Joe Shifted Poisson                 &0.5626   &1.4945    & - &  -1'384.34 \\
                     \hline
		\end{tabular}
	\captionof{table}{Copula families parameters estimates.} \label{table:parm_med}
\end{center}
Table \ref{table:parm_med} shows  that the Joe Geometric copula is the model that best fits the data followed by the Joe Truncated Poisson copula.
}
\subsection{Danish fire insurance data } 
The corresponding data set describes the Danish fire insurance claims collected from the Copenhagen Reinsurance Company for the period 1980-1990. It can be retrieved from  the following website: $www.ma.hw.ac.uk/\sim mcneil/$. This data set has first been considered by Embrechts et al. (1998)   (Example 6.2.9) and explored by Haug et al. (2011).  It consists of three components: loss to buildings, loss to contents and loss to profit. However, in this case, we model the dependence between the first two components. The total number of observations is of 1'501. We only consider the observations where both components are non-null. 
As indicated by the empirical dependence measures in Table \ref{table:dep_measures_Danish}, the level of dependence between these two losses is low.  \\
 \begin{center}
	\begin{tabular}{|c|c|}
  \hline
Pearson's Correlation  &    0.1413   \\
  \hline
Spearman's Rho  &  0.1417  \\
  \hline
Kendall's  Tau &  0.0856	  \\
\hline
Upper tail dependence  & 0.1998    \\
\hline
 	\end{tabular}
\captionof{table}{Dependence measures for the Danish fire insurance. }\label{table:dep_measures_Danish}
\end{center}

The estimation results for each copula is summarized in  Table \ref{table:parm_Danish} below.\\

\begin{center}
	\begin{tabular}{|l|c|c|c|c|}
		\hline
		 Model & $\theta$  & $\alpha$ & $m$&   AIC   \\

		\hline
		Gumbel             &      -  & 1.1762  & - & -133.18  \\
		
		Gumbel Geometric       & 0.9999   & 1.1762 & - & -131.17  \\
		
		Gumbel Truncated Poisson  & 0.0001 &  1.1762 &  - & -131.18 \\
		
		Gumbel Shifted Poisson  &  0.0001  & 1.1762  & - & -131.17  \\
		\hline
                      Frank                  &  -     & 0.8807 & - & -29.12 \\
                    
                      Frank Geometric                 & 0.9999  & 0.8804 & - &-27.12 \\
                     
                      Frank Truncated Poisson                 & 0.0001  & 0.8806 &-  &-27.12 \\
                     
                      Frank Shifted Poisson                 & 0.0001  & 0.8805  & - &-27.12 \\
                     \hline
                     Student                 & -  & 0.1574  & 9.5998 &-47.86 \\
                     
                     Student Geometric        &0.9999 & 0.1576 & 10.0063 & -45.84   \\
                    
                     Student Truncated Poisson     & 0.0001 & 0.1570 & 9.0048 & -45.81    \\       
                     
                     Student Shifted Poisson           &   0.0001& 0.1562 & 8.9833 & -45.42   \\
                     \hline
                     Joe                  &  -     & 1.3585 &- &-204.85 \\
                     
                     Joe Geometric                 & 0.9999  & 1.3585 &-  &-202.83 \\
                     
                     Joe Truncated Poisson                 & 0.0001  & 1.3585 &-  &-202.84\\
                     
                     Joe Shifted Poisson                 &0.0001  & 1.3585 &-  &-202.83 \\
                     \hline
		\end{tabular}
			\captionof{table}{Copula families parameters estimates.} \label{table:parm_Danish}
			\end{center}
It can be seen that the model that best fits the data is the Joe copula followed by the Joe Truncated Poisson copula. The Frank mixture copulas and Student mixture copulas are not a good fit for the data as their AIC is higher by far compared to the Gumbel and Joe mixture copulas families.
\subsection{Reinsurance premiums}
In this section,   we examine the effects of the dependence structure on reinsurance premiums by using the proposed copula models. In practice, it is well known that insurance risks dependency has an impact on  reinsurance. For instance,  Dhaene and Goovaerts (1996)  have shown that  stop loss premium is greater under the dependence assumption than under the independence case. In what follows, we consider the insurance claims data described in  Subsection \ref{subsec:data_VG} where we denote $X$  the loss variable, $Y$ the associated ALAE and $K$ the number of claims for the next accident year. In addition,  two types of reinsurance treaties are analyzed namely: 
\BIT
 \item Excess-of-loss reinsurance, where the claims from $Y_i$ 's are attributed proportionally to the insurer and the reinsurer. For a given observation $(X_i,Y_i)$ the payment for the  reinsurer is described as follows, see Cebrian et al. (2003)
 \BQNY
 			g(X_i,Y_i,r)=\left\{
			 	   \begin{array}{lcl}
         			 0  & \mbox{if} & X_i \leqslant r, \\
         		    X_i-r + \biggl(\frac{X_i- r}{X_i}\biggr) Y_i  & \mbox{if} &  X_i > r
              	\end{array}
      \right.
 \EQNY
leading to a reinsurance premium of the form
   \BQN \label{eq:excessLoss}
   			\kappa(r) = \EE{K}  \EE{g(X_i,Y_i,r)},
   \EQN
   where $r>0$ is the retention \ehe{level}.
   \item Stop loss reinsurance,  where the premium is given by 
	\BQN \label{eq:stopLoss}
			\pi(d)=\EE{ \Bigl(\sum_{i=1}^K (X_i+Y_i) - d \Bigr)_+ }
	\EQN
	and $d$  is a positive deductible.
\EIT
In order to calculate the reinsurance premiums defined above, Monte Carlo  simulations have been implemented. Hereof, 
we assume that $K$ is Poisson distributed with a mean of $156.2$, representing the expected number of claims estimated by the insurance company. Additionally, we use the empirical distributions of $X_i$ and $Y_i$  for the simulation of the claims amount. Regarding the dependence model, the following copulas are considered: independent copula, Joe copula, Geometric Joe Copula, Truncated Poisson Joe copula and the Shifted Poisson Joe copula where the parameters are summarized in Table \ref{table:parm_Vaudoise}.   
 The following steps summarize the implemented  pseudo-algorithm:
\BIT
\item \textbf{Step 1:}  Generate the number of claims $K \sim Poisson(156.2)$.
\item  \textbf{Step 2:} Simulate $(u_{i}, v_{i}), i = 1,\ldots,K$ from the considered copula $C$.
\item  \textbf{Step 3:} Generate the loss and ALAE claims as follows 
									$$(x_i= \hat{F}_{X,n}^{-1}(u_{i}),   y_i= \hat{F}_{Y,n}^{-1}(v_{i})	),  i = 1,\ldots,K,$$
									where $ \hat{F}_{X,n}^{-1}$ and $\hat{F}_{Y,n}^{-1}$ are the inverse of the empirical df of $X$  and $Y$ respectively, with
									$$ \hat{F}_{X,n}(x) = \frac{1}{n} \sum_{i=1} ^ n \mathbf{1}(X_i \leq x), \quad  \hat{F}_{Y,n}(y) = \frac{1}{n} \sum_{i=1} ^ n  \mathbf{1}(Y_i \leq y).$$
\item  \textbf{Step 4:}  Calculate 	the reinsurance premiums $\kappa^b(r)$ 	and $\pi^b(d)$ as in \eqref{eq:excessLoss} and \eqref{eq:stopLoss} respectively.	
\item   \textbf{Step 5:}	 \textbf{Step 1}	-\textbf{Step 4} are repeated $B$	times and the estimators of the reinsurance premiums are given by 
$$ \hat{\kappa}(r) = \frac{1}{B} \sum_{b=1} ^B \kappa^b(r), \quad \hat{\pi}(d) = \frac{1}{B} \sum_{b=1} ^B \pi^b(d).$$
\EIT
The estimation results presented in Table \ref{table:reins_prem} are obtained from repeating  \textbf{Step 1}	-\textbf{Step 4} 100'000 times. These amounts are expressed in CHF million. 
\begin{center}
	\begin{tabular}{|l||c|c|c||c| c|c||}
		\hline
			 & 	 \multicolumn{3}{c||}{ $\hat{\kappa}(r)$  } &\multicolumn{3}{c||}{ $\hat{\pi}(d)$ }  \\
		\hline
			Copula model		 & $r=1$ & $r=5$& $r=10$ & $d=10$& $d=20$& $d=30$\\
		\hline
					Independent & 13.1137  & 6.5692 & 3.0971  & 14.7530& 7.5145& 3.5738 \\
		
                     \hline
                     Joe         & 13.6950 &  6.7776  &3.2396  & 15.1056& 7.7691& 3.8233\\        
                     \hline
                     Joe Geometric                & 13.4483  &6.7365 &3.1619 &14.8975&7.6797&3.7177 \\
                     \hline
                     Joe Truncated Poisson     &  13.4038 & 6.7183 & 3.0929 &14.8016 & 7.6698&  3.6493\\
                     \hline
                     Joe Shifted Poisson             &13.4776  &6.6789 & 3.1081&14.9250& 7.6266& 3.6702\\
                     \hline
		\end{tabular}
			\captionof{table}{Reinsurance premiums  with respect to copula models.} \label{table:reins_prem}
			\end{center}
Table 13 shows that the reinsurance premiums $\hat{\kappa}(r)$ and $\hat{\pi}(d)$ are lower under the independence hypothesis. Hence, the portfolio is less risky when the loss variable $X_i$ and the ALAE variable $Y_i$ are assumed to be independent. 
Furthermore,  when the retention limit $r$ increases for the excess of loss treaty,  the reinsurance premiums estimates $\hat{\kappa}(r)$ under the Joe mixture copula models tend to the estimated values under the independence assumption. Conversely, for the stop loss treaty,  the higher the deductible $d$  the higher the deviation from the independence hypothesis. \\
Furthermore, by comparing the results for each copula model, it can be seen that the Joe copula generates the highest reinsurance premiums. This result is expected given that the strongest dependence structure is obtained under the Joe copula. On the other hand, the weakest dependence model for this data is observed under the Joe truncated Poisson copula as the reisurance premiums $\hat{\kappa}(r)$ and $\hat{\pi}(d)$ are the smallest for different values of  $r$ and $d$.
\COM{
\begin{center}
	\begin{tabular}{|l||c|c|c||c| c|c||}
		\hline
			 & 	 \multicolumn{3}{c||}{ $\kappa(r)$  } &\multicolumn{3}{c||}{ $\pi(d)$ }  \\
		\hline
			Copula models		 & $r=1$ & $r=5$& $r=10$ & $d=10$& $d=20$& $d=30$\\
		\hline
					Independence & 13113'681  & 6'569'242 & 3097074 & 14'752'970& 7'514'458& 3573805\\
		
                     \hline
                     Joe         & 13'694'958 &  6'777'595 &3239632  & 15'105'628& 7'769'147& 3823279\\        
                     \hline
                     Joe Geometric                & 13'448'312  &6544036 &3049173 &14'897'486&7414603&3614379 \\
                     \hline
                     Joe Truncated Poisson                 &  13'212'492 & 6718298 & 3046561 &14'625365& 7669821&  3573074\\
                     \hline
                     Joe Shifted Poisson                 &13'477'576  &6678911 & 3108136&14'924'985& 7626605& 3670225\\
                     \hline
		\end{tabular}
			\captionof{table}{Copula families parameters estimates.} \label{table:parm_Danish}
			\end{center}

}

\COM{
\section{Conclusion} 

In this paper, we propose a new family set of mixture copulas for large bivariate risks. 
This idea was first inspired by Zhang et al. (2015)   with the use of the Geometric distribution to construct a new set of Geometric copulas. \\
Therefore, let $(X_i,Y_i)$ model the $i^{th}$ claim of both portfolios and let $N$ be the total number of such claims. Let $\Lambda=N|N \ge 1$. When $N\ge1$, we denote $(X_{N:N},Y_{N:N})$ the maximal claims amount in both portfolios with distribution function $C$. Following Zhang et al. (2015), we develop a general expression of the new mixture copula distribution function $C$ associated to the maximal claims amount and derive the general form of the probability density function $c$ of $C$. $C$ depends on the original copula $Q$ which models the dependence between the risks $X_i$ and $Y_i$. For application, we introduce the Schifted Poisson and Truncated Poisson distributions to model the number of claims in both portfolios.
Also, we consider the Gumbel, Frank, Joe and Student copulas with $Q$ being their copulas distribution function to construct the new family of copulas. To see the different application of the new mixture copulas, four insurance datasets are studied and in each dataset we assess the best fit of the new mixture copulas families by using the AIC \ehe{criteria}.
Based on the pseudo-maximum likelihood method, the parameters of $Q$ and $C$, $\alpha$ and $\theta$ respectively, are estimated by maximizing the pseudo log-likelihood function.\\

This paper gives new insights on the modelling of large claims where the number of claims is a random variable following a certain distribution. In practice, the number of claims $N$ generally follows a Poisson distribution. It can be seen that in some of the cases the model that best fits the data is the mixture copula. For e.g. the data based on the general liability losses and their associated ALAE, the Joe \ehe{Shifted} Poisson copula and the Gumbel Schifted Poisson copula outperforms the Joe and Gumbel copula respectively. An idea to develop in a 
future paper is how to generalize the bivariate mixture copula to a multivariate mixture copula.

}

\section{Appendix} 
\subsection{Proofs}
 
Derivation of \eqref{eq:ShiftedPoissonpdf}-\eqref{eq: TruncatedPoisPdf}: We show first \eqref{eq:Copula_PDFF}. 
The corresponding joint  density $c$ of the df $C$  is given by 
\BQN \label{eq:pdf_def}
c(u_1,u_2) =\frac{\partial C (u_1,u_2)}{\partial u_1 \partial u_2}
=\frac{\partial L_\Lambda (- \ln Q_\alpha(v_1,v_2))}{\partial u_1 \partial u_2},
\EQN
where 
$$C(v_1,v_2)=L_\Lambda (- \ln Q_\alpha(v_1,v_2)), \quad \green{v_i= e^{-L^{-1}_\Lambda(u_i)}}, \quad i=1,2.$$
In view of \eqref{eq:pdf_def}, the partial derivative of $C$ with respect to $u_1$ is 
$$\frac{\partial L_\Lambda (-\ln Q_\alpha(v_1,v_2))}{\partial u_1}= \frac {1}{Q_\alpha(v_1,v_2)} L'_\Lambda (- \ln Q_\alpha(v_1,v_2))\frac{-\partial Q_\alpha(v_1,v_2)}{\partial v_1} \frac{\partial v_1}{\partial u_1}$$
leading to
\BQNY
c(u_1,u_2)
&=&
\frac{\partial}{\partial v_2}
\Biggl(
\frac{\partial L_\Lambda (-\ln Q_\alpha(v_1,v_2))}{\partial u_1}
\Biggr)
 \frac{\partial v_1}{\partial u_1} \frac{\partial v_2}{\partial u_2}\\
&=&  \frac{\partial v_2}{\partial u_2}
\Biggl(
L''_\Lambda(-\ln Q_\alpha(v_1,v_2))\frac{\frac{\partial Q_\alpha(v_1,v_2)}{\partial v_1}\frac{\partial Q_\alpha(v_1,v_2)}{\partial v_2}}{Q_\alpha^2(v_1,v_2)} \\
&&+ 
L'_\Lambda(-\ln Q_\alpha(v_1,v_2))\frac{\frac{-\partial^2 Q_\alpha(v_1,v_2)}{\partial v_1 \partial v_2}Q_\alpha(v_1,v_2) +\frac{\partial Q_\alpha(v_1,v_2)}{\partial v_1}\frac{\partial Q_\alpha(v_1,v_2)}{\partial v_2}}{Q_\alpha^2(v_1,v_2)}
\Biggr)\\
&=&
 \frac{\frac{\partial v_1}{\partial u_1} \frac{\partial v_2}{\partial u_2}}{Q_\alpha^2(v_1,v_2)}
 \Biggl(
 \Bigl(L''_\Lambda(-\ln Q_\alpha(v_1,v_2))+L'_\Lambda(-\ln Q_\alpha(v_1,v_2))\Bigr)\frac{\partial Q_\alpha(v_1,v_2)}{\partial v_1}\frac{\partial Q_\alpha(v_1,v_2)}{\partial v_2}\\
&& -
 L'_\Lambda(-\ln Q_\alpha(v_1,v_2))Q_\alpha(v_1,v_2)\frac{\partial^2 Q_\alpha(v_1,v_2)}{\partial v_1 \partial v_2}
\Biggr).
\EQNY
\\
We derive next the pdf $c_\Theta$ in \eqref{eq:ShiftedPoissonpdf}: In this case, $\Lambda$ follows a shifted Poisson distribution. In view of \eqref{eq:Copula_PDFF},  we need to  compute at first the following components:
\BQNY 
L'_\Lambda(-\ln Q_\alpha(v_1,v_2)) &=& -e^{-\theta(1-Q_\alpha(v_1,v_2))}Q_\alpha(v_1,v_2)(1+\theta Q_\alpha(v_1,v_2)), \\
L''_\Lambda(-\ln Q_\alpha(v_1,v_2))&=& e^{-\theta(1-Q_\alpha(v_1,v_2))}Q_\alpha(v_1,v_2)(1+3\theta Q_\alpha(v_1,v_2)+\theta^2 Q_\alpha ^2(v_1,v_2)),
\EQNY
where for $i=1,2$, $v_i=e^{-L^{-1}_\Lambda(u_i)}$  which implies $ u_i= v_i e^{-\theta (1-v_i)}$
and thus 
$\frac{\partial v_i}{\partial u_i}= \frac{e^{-\theta(1-v_i)}}{1+\theta v_i}.$
By replacing these components  into \eqref{eq:Copula_PDFF}, we have
\ehe{\BQNY
c_\Theta(u_1,u_2)
&=& 
\frac{1}{Q_\alpha(v_1,v_2)^2}\frac{e^{\theta(2-v_1-v_2)}}{(1+\theta v_1)(1+\theta v_2)} 
\Biggl[
\Bigl(
e^{-\theta (1-Q_\alpha(v_1,v_2))} Q_\alpha(v_1,v_2)(1+3\theta Q_\alpha(v_1,v_2)+\theta^2 Q_\alpha ^2(v_1,v_2))\\
&&
-e^{-\theta (1-Q_\alpha(v_1,v_2))}Q_\alpha(v_1,v_2)(1+\theta Q_\alpha(v_1,v_2)
\Bigr)
\frac{\partial Q_\alpha(v_1,v_2)}{\partial v_1}\frac{\partial Q_\alpha(v_1,v_2)}{\partial v_2}\\
&&
+ e^{-\theta (1-Q_\alpha(v_1,v_2))}Q_\alpha (v_1,v_2)(1+\theta Q_\alpha(v_1,v_2)) Q_\alpha(v_1,v_2)\frac{\partial^2 Q_\alpha(v_1,v_2)}{\partial v_1 \partial v_2}
\Biggr]\\
&=&\frac{1}{Q_\alpha(v_1,v_2)^2}\frac{e^{\theta(2-v_1-v_2)}}{(1+\theta v_1)(1+\theta v_2)}  \times \\
&& \Biggl[
e^{-\theta (1-Q_\alpha(v_1,v_2))}Q_\alpha (v_1,v_2)\Bigl(1+3\theta Q_\alpha(v_1,v_2)+\theta^2 Q_\alpha ^2(v_1,v_2)-1-\theta Q_\alpha(v_1,v_2)\Bigr)\frac{\partial Q_\alpha(v_1,v_2)}{\partial v_1} \frac{\partial Q_\alpha(v_1,v_2)}{\partial v_2}\\
&& +e^{-\theta (1-Q_\alpha(v_1,v_2))}Q_{\alpha}^2 (v_1,v_2)(1+\theta Q_\alpha(v_1,v_2)) \frac{\partial^2 Q_\alpha(v_1,v_2)}{\partial v_1 \partial v_2}
\Biggr]\\
&=&\frac{1}{Q(v_1,v_2)^2}\frac{e^{\theta(2-v_1-v_2)}}{(1+\theta v_1)(1+\theta v_2)}  \times \\
&& \Biggl[
e^{-\theta (1-Q_\alpha(v_1,v_2))}Q_\alpha (v_1,v_2)\Bigl(2\theta Q_\alpha(v_1,v_2)+\theta^2 Q_\alpha ^2(v_1,v_2)\Bigr)\frac{\partial Q_\alpha(v_1,v_2)}{\partial v_1} \frac{\partial Q_\alpha(v_1,v_2)}{\partial v_2}\\
&& +e^{-\theta (1-Q_\alpha(v_1,v_2))}Q_{\alpha}^2 (v_1,v_2)(1+\theta Q_\alpha(v_1,v_2)) \frac{\partial^2 Q_\alpha(v_1,v_2)}{\partial v_1 \partial v_2}
\Biggr]\\
&=&\frac{1}{Q(v_1,v_2)^2}\frac{e^{\theta(2-v_1-v_2)}}{(1+\theta v_1)(1+\theta v_2)}  \times \\
&& \Biggl[
e^{-\theta (1-Q_\alpha(v_1,v_2))}Q_{\alpha}^2 (v_1,v_2)\Bigl(2\theta +\theta^2 Q_\alpha(v_1,v_2)\Bigr)\frac{\partial Q_\alpha(v_1,v_2)}{\partial v_1} \frac{\partial Q_\alpha(v_1,v_2)}{\partial v_2}\\
&&+e^{-\theta (1-Q_\alpha(v_1,v_2))}Q_{\alpha}^2 (v_1,v_2)(1+\theta Q_\alpha(v_1,v_2)) \frac{\partial^2 Q_\alpha(v_1,v_2)}{\partial v_1 \partial v_2}
\Biggr]\\
&=&\frac{1}{Q(v_1,v_2)^2}\frac{e^{\theta(2-v_1-v_2)}}{(1+\theta v_1)(1+\theta v_2)}e^{-\theta (1-Q_\alpha(v_1,v_2))}Q_{\alpha}^2 (v_1,v_2)  \times \\
&& \Biggl[
 \theta \frac{\partial Q_\alpha(v_1,v_2)}{\partial v_1} \frac{\partial Q_\alpha(v_1,v_2)}{\partial v_2} \Bigl(2 +\theta Q_\alpha(v_1,v_2)\Bigr)+(1+\theta Q_\alpha(v_1,v_2)) \frac{\partial^2 Q_\alpha(v_1,v_2)}{\partial v_1 \partial v_2}
\Biggr]\\
&=&\frac{e^{\theta(2-v_1-v_2)} e^{\theta(Q_\alpha(v_1,v_2)-1)} }{(1+\theta v_1)(1+\theta v_2)} 
\Biggl[
 \frac{\partial^2 Q_\alpha(v_1,v_2)}{\partial v_1 \partial v_2}(1+\theta Q_\alpha(v_1,v_2))+\theta \frac{\partial Q_\alpha(v_1,v_2)}{\partial v_1} \frac{\partial Q_\alpha(v_1,v_2)}{\partial v_2} (2+\theta Q_\alpha(v_1,v_2))
 \Biggr].
\EQNY}
\\
Next, we show \eqref{eq: TruncatedPoisPdf}: Since $\Lambda$ follows a truncated Poisson distribution,  in light of \eqref{eq:Copula_PDFF}, the joint density $c_\Theta$ is expressed in terms of (set $\eta_\theta= e^{-\theta}/(1-e^{-\theta})$)
\BQNY
L'_\Lambda(-\ln Q_\alpha(v_1,v_2)) &=&-\ehe{\eta_\theta}\theta Q_\alpha(v_1,v_2) e^{\theta Q_\alpha(v_1,v_2)},\\
L''_\Lambda (-\ln Q_\alpha(v_1,v_2))&=&\ehe{\eta_\theta}\theta Q_\alpha(v_1,v_2) e^{\theta Q_\alpha(v_1,v_2)} (1+\theta Q_\alpha(v_1,v_2)),
\EQNY
where  for $i=1,2$,  $v_i=e^{-L^{-1}_\Lambda(u_i)}$ and $ u_i=\frac{e^{-\theta}}{1-e^{-\theta}}(v_i^\theta-1)$
with $\frac{\partial v_i}{\partial u_i}=\frac{1-e^{-\theta}}{\theta}e^{\theta(1-v_i)}.$
By substituting the above components in the joint density expressed in \eqref{eq:Copula_PDFF}, we obtain
\BQNY
c_\Theta(u_1,u_2)
&=& 
\Biggl(\frac{1-e^{-\theta}}{\theta }\Biggr)^2  
 \frac{e^{\theta(2-v_1-v_2)}}{\ehe{Q_\alpha^2}(v_1,v_2)} 
 \Biggl[\Biggl(\ehe{\eta_\theta}\theta Q_\alpha(v_1,v_2) e^{\theta Q(v_1,v_2)} (1+\theta Q_\alpha(v_1,v_2))\\
&& -\ehe{\eta_\theta} \theta Q_\alpha(v_1,v_2) e^{\theta Q_\alpha(v_1,v_2)}\Biggr)\frac{\partial Q_\alpha(v_1,v_2)}{\partial v_1}\frac{\partial Q_\alpha(v_1,v_2)}{\partial v_2}\\
&&+\ehe{\eta_\theta} \theta Q_\alpha(v_1,v_2) e^{\theta Q_\alpha(v_1,v_2)}Q_\alpha(v_1,v_2)\frac{\partial^2 Q_\alpha(v_1,v_2)}{\partial v_1 \partial v_2} \Biggr]\\
&=&(1-e^{-\theta}) e^{\theta[ 1-v_1-v_2 + Q_\alpha(v_1,v_2)]}\Biggl( \frac{\partial Q_\alpha(v_1,v_2)}{\partial v_1} \frac{\partial Q_\alpha(v_1,v_2)}{\partial v_2}+\ehe{\frac 1 {\theta}}\frac{\partial^2 Q_\alpha(v_1,v_2)}{\partial v_1 \partial v_2}\Biggr).
\EQNY

\QED
\proofprop{tauHok}
Since $G$ has \FRE marginals, by assumption \eqref{copA}, we have that 
$$ \limit{n} G^n(nx,ny)= \mathcal{G}(x,y), \quad x,y\in (0,\IF),$$
where $\mathcal{G}$ has copula $Q_{\ehe{A}}$ and thus $\tau(\mathcal{G})=\tau (Q_{{A}})$.
We have thus with $F_n(x,y)= \EE{G^{\Lambda_n}(x,y)}$ using further \eqref{Z}
\bqn{
\label{XX} \limit{n}F_n(nx_n,n y_n)=  \limit{n}\EE{G^{n \frac{\Lambda_n}n}(nx_n,ny_n)}= \mathcal{G}(x,y), \quad x,y\in (0,\IF)
}
for any $x_n,y_n$ such that $\limit{n} x_n= x$ and $\limit{n} y_n= y$. Consequently, 
\bqny{
\tau(C_n)&=& 4 \int_{(0,\IF)^2} F_n(x,y) \, d F_n(x,y)- 1\\
&=& 4 \int_{(0,\IF)^2} F_n(nx,ny) \, d F_n(nx,ny)- 1\\
&\to& 4 \int_{(0,\IF)^2} \mathcal{G}(x,y) \, d  \mathcal{G}(x,y)- 1, \quad n\to \IF\\
&=& \tau(  \mathcal{G}),
}
where the convergence above follows by Lemma 4.2 in Hashorva (2007)  (see also Resnick and Zeber (2013)  and Kulik and Soulier (2015) for more general results). Next, the convergence in \eqref{XX} implies 
\bqny{  
\limit{n }F_{ni}(ns_n)=  \limit{n}\EE{G^{n \frac{\Lambda_n}n}_i(ns_n)}= \mathcal{G}_i(s), \quad s\in (0,\IF), i=1,2
}
for any $s_n,n\ge 1$ such that $\limit{n} s_n = s$, where $F_{ni},G_i, \mathcal{G}_i$ is the $i$th marginal df of $F_n, G$, and 
$\mathcal{G}$, respectively. Hence, with similar arguments as above, we have 
\bqny{
\rho_S(C_n)&=& 12 \int_{(0,\IF)^2} F_n(x,y) \, d F_{n1}(x) d F_{n2}(y)- 3\\
&=& 12 \int_{(0,\IF)^2} F_n(nx,ny) \, d F_{n1}(nx) d F_{n2}(nx)  - 3\\
&\to& 12 \int_{(0,\IF)^2} \mathcal{G}(x,y) \, d  \mathcal{G}_1(x)d  \mathcal{G}_2(x)- 3, \quad n\to \IF\\
&=& \rho_S(  \mathcal{G})
 }
establishing the proof.
\QED

\proofprop{prop1} 
For $ v= e^{- L_\Lambda^{-1}(1-u)}$ we have 
$$ 1-  L_\Lambda(- \ln v) \sim   u, \quad u\downarrow 0, \quad \lim_{u \downarrow0 } v=1.$$
By the assumption that $\EE{\Lambda}$ is finite we have 
\bqn{\label{LF}
 1- L_\Lambda(t) \sim -L_\Lambda'(0) t= \EE{\Lambda} t, \quad t\to 0.
 }
Since further 
$$ 
\mu_Q= 2- \lim_{u\downarrow 0} \frac{ Q(1- u,1-u)}{u} = 2- \lim_{v\uparrow 1} \frac{\ln Q(v,v)}{\ln v}
$$
and $ \lim_{v\uparrow 1} Q(v,v)=1,$ then using \eqref{CopulaF} and \eqref{LF} we obtain 
\BQNY
\mu_C &=&2- \lim_{u \downarrow 0} u^{-1}[1- C(1- u, 1-u))]\\
&=&2 - \lim_{u \downarrow 0} u^{-1}\Bigl [1-  L_\Lambda\bigl(-\ln Q(v ,v)\bigr) \Bigr]\\
&=&2- \lim_{u \downarrow 0} \frac{ 1-  L_\Lambda\bigl(-\ln Q(v ,v)\bigr) }{1- L_\Lambda(- \ln v)}\\
&=&2- \lim_{v \uparrow 1} \frac{ \ln Q(v ,v) }{\ln v } \\
&=&2- [2- \mu_Q]=\mu_Q,
\EQNY
hence the first claim follows.  Next, in view of \eqref{md1} we have 
\BQNY
\limit{n} n [1- G(n x, ny)]= - \ln H(x,y), \quad x,y\in (0,\IF),
\EQNY
hence as $n\to \IF$
\BQNY
n [1- G(n x, ny)] \sim \frac{ 1- G(n x,ny)}{1- G(n, n)}  \sim  - \ln H(x,y), \quad x,y \in (0,\IF).
\EQNY
Let $a_n,n\ge 1$ be non-negative constants such that $\limit{n} a_n= \IF$. 
\COM{By the above (set $q_n:= 1- G(a_n,a_n))$
\BQN\label{12}
\frac{ 1- G(a_n x, a_n y)}{q_n }  \sim  - \ln H(x,y), \quad x,y \in (0,\IF), \quad  
q_n  \sim - a_n^{-1}\ln H(1,1).
\EQN
Note that since $\limit{n}q_n=0$, 
 then $- \ln G(a_n,a_n) \sim q_n$ as $n\to \IF$.} By the above and \eqref{LF} 
\BQNY
n[ 1- F(a_nx,a_ny)] &=& n[1- L_\Lambda(- \ln G(a_nx,a_ny))] \sim \EE{\Lambda} n(- \ln G(a_nx,a_ny))
\EQNY
as $n\to \IF$. Setting now $a_n= \EE{\Lambda} n$ we have thus 
as $n\to \IF$
\BQNY 
 n [1- F(a_n x, a_n y)]  &\sim&  a_n \frac{ 1- F(a_nx, a_n y)}{\EE{\Lambda}} \\
&= &    a_n \frac{1- L_\Lambda( - \ln G(a_n x, a_n y)) }{\EE{\Lambda}} \\
&\sim &   a_n (- \ln G(a_n x, a_n y))  \\
&\sim &   a_n [1- G(a_n x, a_n y)[  \\
&\sim &   \EE{\Lambda}\Bigl(- \ln H( x \EE{\Lambda},y\EE{\Lambda})\Bigr)\\
&=&  -\ln H( x ,y)
 \EQNY
establishing the proof. \QED


For our study, we consider several copula families for $Q_{\alpha}$, which are described hereafter.
\subsection{ Gumbel Copula}
The df of a Gumbel copula with a dependence parameter $\alpha \ehe{\ge} 1$ is given by 
\BQNY
Q_{\alpha}(v_1,v_2)= \exp\biggl( -\Bigl((-\ln v_1)^{\alpha}+(-\ln v_2)^{\alpha}\Bigr)^\frac{1}{\alpha} \biggl)
\EQNY
by \ebe{differentiating}  $ Q_{\alpha}(v_1,v_2)$ with respect to $v_1$ we have
\BQN
\frac{\partial Q_{\alpha}(v_1,v_2)}{\partial v_1}= \frac{1}{v_1} (-\ln v_1)^{\alpha -1}\Biggl((-\ln v_1)^{\alpha}+(-\ln v_2)^{\alpha}\Biggr)^\frac{1}{\alpha -1}e^{-\Bigl((-\ln v_1)^{\alpha}+(-\ln v_2)^{\alpha}\Bigr)^\frac{1}{\alpha}},\notag
\EQN
and the corresponding joint density is expressed as follows
\BQNY
q_{\alpha}(v_1,v_2)
&=& \frac{(-\ln v_1)^{\alpha -1} (-\ln v_2)^{\alpha - 1}}{v_1 v_2}  \Bigl(a^{\frac{2}{\alpha} -2} + (\alpha -1)a^{\frac{1}{\alpha} -2}\Bigr) e^{-a^\frac{1}{\alpha}},
\EQNY
where $a= (-\ln v_1)^{\alpha}+(-\ln v_2)^{\alpha}$.

\subsection{Frank Copula} The df of a Frank copula with a dependence parameter $ \alpha \ne 0$ is of the form
\BQN
Q_{\alpha}(v_1,v_2)=\frac{-1}{\alpha}\ln \Bigl(1+ \frac{(e^{-\alpha v_1}-1)(e^{-\alpha v_2}-1)}{e^{-\alpha}-1}\Bigr),\notag
\EQN
which yields the \ebe{partial derivative} of $Q_{\alpha}(v_1,v_2)$  with respect to $v_1$ as follows
\BQNY
\frac{\partial Q_{\alpha}(v_1,v_2)}{\partial v_1}=\frac{ e^{-\alpha v_1}(e^{-\alpha v_2}-1)}{(e^{-\alpha}-1)+(e^{-\alpha v_1}-1)(e^{-\alpha v_2}-1)} \notag
\EQNY
and the associated \ebe{pdf} is given by
\BQNY
q_{\alpha}(v_1,v_2)
&=& \frac{\alpha (1- e^{-\alpha})e^{-\alpha(v_1+v_2)}}{\Bigl((1- e^{-\alpha})-(1-e^{-\alpha v_1})(1-e^{-\alpha v_2})\Bigr)^2}.
\EQNY

\subsection{Joe copula}
The Joe copula with  dependence parameter $\alpha \ehe{\ge} 1$  has  df
\BQN
Q_{\alpha}(v_1,v_2)=1-\Bigl((1-v_1)^{\alpha}+(1-v_2)^{\alpha}-(1-v_1)^{\alpha}(1-v_2)^{\alpha}\Bigr)^{\frac{1}{\alpha}}.\notag
\EQN
Deriving $Q_{\alpha}(v_1,v_2)$ with respect to $v_1$ we obtain
\BQN
\frac{\partial Q_{\alpha}(v_1,v_2)}{\partial v_1}=(1-v_1)^{\alpha -1}(1-(1-v_2)^{\alpha})\Bigl((1-v_1)^{\alpha}+(1-v_2)^{\alpha}-(1-v_1)^{\alpha}(1-v_2)^{\alpha}\Bigr)^{\frac{1}{\alpha}-1}.\notag
\EQN
The associated \ebe{pdf} is obtained by \ehe{differentiating}   $Q_{\alpha}(v_1,v_2)$ with respect to $v_1$ and $v_2$ leading to 
\BQNY
q_{\alpha}(v_1,v_2)
&=& (1-v_1)^{\alpha -1}(1-v_2)^{\alpha-1}\Bigl(\alpha -1 +b\Bigr)b^{\frac{1}{\alpha}-2	},
\EQNY
where $b=(1-v_2)^{\alpha}-(1-v_1)^{\alpha}(1-v_2)^{\alpha}$.

\subsection{Student Copula}
\ehe{Let  $t_m$ be the df of a Student random variable with degree of freedom $m$ and write $t_m^{-1} $ for its inverse.}
The df of the Student copula, with correlation $\alpha \in \ehe{(-1,1)}$ and degree of freedom $m>0$  can be expressed as follows
\BQNY
Q_{\alpha, m}(v_1,v_2)
&=& t_{\alpha ,m} (t_{m}^{-1}(v_1),t_{m}^{-1}(v_2))\\
&=&\int_{- \infty}^{t_{m}^{-1}(v_1)}\int_{- \infty}^{t_{m}^{-1}(v_2)}\frac{1}{\sqrt{2 \pi (1-\alpha ^2)}}\Bigl(1+ \frac{s^2-2 \alpha st + t^2}{m (1-\alpha ^2)}\Bigr)^{-(m+2)/2} ds dt .
\EQNY
Its \ebe{partial} derivative with respect to $v_1$ is given by
\BQNY
\frac{\partial Q_{\alpha}(v_1,v_2)}{\partial v_1}=t_{m+1}\Biggl(\frac{t^{-1}_{m}(v_2)-\alpha t^{-1}_{m}(v_1)}{\sqrt{\frac{(m+(t^{-1}_{m}(v_1))^2)(1-\alpha ^2)}{m+1}}}\Biggr),
\EQNY
whereas the corresponding \ebe{pdf} is  
\BQNY
q_{\alpha ,m}(v_1,v_2)
&=& \frac{1}{2 \pi \sqrt{1-\alpha ^2}}\frac{1}{k(t^{-1}_{m}(v_1)) k(t^{-1}_{m}(v_2))}\Bigl(1+\frac{{t^{-1}_{m}(v_1)}^2+{t^{-1}_{m}(v_2)}^2 - 2 \alpha t^{-1}_{m}(v_1) t^{-1}_{m}(v_2)}{m(1-\alpha ^2)}\Bigr)^{-\frac{m+2}{2}},
\EQNY
where for $ i=1,2$
$$k (t^{-1}_{m}(v_i))=\frac{\Gamma(\frac{m+1}{2})}{\Gamma(\frac{m}{2}) \sqrt{\pi m}}\Bigl(1+\frac{t^{-1}_{m}(v_i)^2}{m}\Bigr)^{-\frac{m+1}{2}}. $$ 
\\


\textbf{Acknowledgments}. Thanks to reviewers and the Editor for several suggestions. 
E. Hashorva is partially supported by the Swiss National Science Foundation grants 200021-13478, 200021-140633/1. 
G. Ratovomirija is partially supported by the project RARE -318984
 (an FP7  Marie Curie IRSES Fellowship) and Vaudoise Assurances.

\COM{ \bibliography{BiB}}
\end{document}